\documentclass[a4paper,11pt,oneside]{article}

\usepackage[english]{babel}
\usepackage[T1]{fontenc}
\usepackage[utf8]{inputenc}

\usepackage[pdftex,unicode]{hyperref}
\hypersetup{pdftitle=Cultural Diversity and Its Impact on Governance}
\hypersetup{pdfauthor=Tomáš Evan and Vladimír Holý}

\usepackage{color}
\definecolor{mycolor}{rgb}{0.60,0.30,0.00}
\hypersetup{colorlinks=true, linkcolor=mycolor, anchorcolor=mycolor, citecolor=mycolor, filecolor=mycolor, urlcolor=mycolor}
\definecolor{mygray}{gray}{0.65}
\newcommand{\GC}[1]{\textcolor{mygray}{#1}}

\usepackage[margin=60pt]{geometry}
\usepackage[authoryear]{natbib}

\usepackage{amsmath}
\usepackage{amssymb}
\usepackage{graphicx}
\usepackage[group-minimum-digits=3]{siunitx}
\usepackage{booktabs}

\usepackage{longtable}
\usepackage{pdflscape}
\usepackage{caption}

\begin{document}

\begin{center}
{\Large \bfseries Cultural Diversity and Its Impact on Governance}
\end{center}

\begin{center}
{\bfseries Tomáš Evan} \\
Czech Technical University in Prague \\
Thákurova 2077/7, 160 00 Prague 6, Czech Republic \\
\href{mailto:tomas.evan@fit.cvut.cz}{tomas.evan@fit.cvut.cz}
\end{center}

\begin{center}
{\bfseries Vladimír Holý} \\
Prague University of Economics and Business \\
Winston Churchill Square 4, 130 67 Prague 3, Czech Republic \\
\href{mailto:vladimir.holy@vse.cz}{vladimir.holy@vse.cz} \\
\end{center}

\noindent
\textbf{Abstract:}
Hofstede's six cultural dimensions make it possible to measure the culture of countries but are criticized for assuming the homogeneity of each country. In this paper, we propose two measures based on Hofstede's cultural dimensions which take into account the heterogeneous structure of citizens with respect to their countries of origin. Using these improved measures, we study the influence of heterogeneous culture and cultural diversity on the quality of institutions measured by the Worldwide Governance Indicators, which encompass Voice and Accountability, Political Stability, Government Effectiveness, Regulatory Quality, Rule of Law, and Control of Corruption. We use a linear regression model allowing for dependence in spatial and temporal dimensions as well as high correlation between the governance indicators. Our results show that the effect of cultural diversity improves some of the governance indicators while worsening others depending on the individual Hofstede cultural dimension. Most notably, diversity in the Power Distance cultural dimension has a negative impact on Political Stability.
\\

\\

\\

\section{Introduction}
\label{sec:intro}

The importance of institutions in the development of countries has been accepted for some time now. Countries' institutions and policies are based on the convictions and reasoning of the holders of the culture which is passed on from generation to generation. It is therefore important to find the mechanisms by which culture influences the institutions and well-being of its holders. Therefore, it is as important as it is interesting to put culture into the picture and to find the mechanisms through which culture influences both the institutions and well-being of its holders. Even the terms culture and institutions are intertwined. As we want to describe the influence of one on the other, we need to distinguish between the two. Thus, we will define both relatively narrowly. Culture is defined as informal while institutions refer to formalized behaviour in line with \cite{North1990} and \cite{Olson1996}, i.a. This allows us to use Hofstede's cultural dimensions \citep{Hofstede2010}, namely Power Distance, Individualism, Masculinity, Uncertainty Avoidance, Long-Term Orientation, and Indulgence, as a proxy informal culture. In addition, as a proxy for formal culture, we employ the Worldwide Governance Indicators \cite{Kaufmann2011}, namely Voice and Accountability, Political Stability, Government Effectiveness, Regulatory Quality, Rule of Law, and Control of Corruption. Our paper has the following two goals:

First, we propose two cultural measures based on Hofstede's cultural dimensions which take into account the heterogeneity of the culture within a country. By incorporating the culture of the migrating populations into these measures, we mitigate some of the criticism of Hofstede's cultural dimensions. For each of Hofstede's cultural dimensions, we define the cultural level indicator as the average of the cultural dimensions assigned to each citizen based on the country of origin. The interactions of cultural minorities with the prevalent culture might have a distinct effect on governance beyond the change in the average of the country's cultural dimensions. Therefore, for each of Hofstede's cultural dimensions, we also determine the cultural diversity indicator defined as the standard deviation of the cultural dimension over all citizens of a country with respect to their different origins.

Second, using the proposed indicators, we investigate the influence of culture and cultural diversity on the quality of governance of countries. In this, we follow our previous research \cite{Holy2022d} in which we study the dependence of the six Worldwide Governance Indicators on Hofstede's six cultural dimensions. In the current paper, we improve the regression model by relaxing the strong assumption of homogeneity of culture in each country. We let the culture be heterogonous with respect to the different countries of origin of the migrating populations. This allows us to better describe the complex relationship between culture and the quality of institutions and fill the gap in the literature on this difficult to measure subject.

The rest of the paper is structured as follows. In Section \ref{sec:theory}, we review the theoretical background of the influence of migration, culture, and cultural diversity on quality of governance. In Section \ref{sec:measure}, we describe the Worldwide Governance Indicators, the Hofstede's cultural dimensions, and various cultural diversity measures. In Section \ref{sec:model}, we propose our novel measures of heterogeneous culture and cultural diversity and use them in a regression model of governance. In Section \ref{sec:est}, we conduct an empirical study assessing the impact of culture and cultural diversity on governance in 115 countries from 2000 to 2020. In Section \ref{sec:disc}, we discuss limitations of our approach and possible directions for future research. We conclude the paper in Section \ref{sec:con}.

\section{Theoretical Background}
\label{sec:theory}

\subsection{Migration as a Source of Cultural Diversity}
\label{sec:theoryMigration}

Since time immemorial large migrations were considered cataclysmic events. Neolithic tribes before the discovery and proliferation of the institution of private property and thus agriculture \citep{North1977} operated under common ownership, which led to the exhaustion of local resources. This routine and arguably inevitable overhunting and overfishing was caused by the principle described by William Forster Lloyd already in 1833 as the tragedy of the commons \citep{Lloyd1980}. The tribe was forced to move in order to survive, which led to violent clashes with other tribes in its path and very often with the annihilation or mass enslavement of the weaker side. There is only a handful of examples in history where a quasi-peaceful solution was found between different ``cultures'' sharing the same land. Eventually, migrations of entire tribes were replaced by migrations of specific groups or individuals usually too aggressive or otherwise ill-fitting so as to be exiled or motivated to leave. Instead of endangering entire domestic populations these smaller groups either expanded “civilisation” on the frontiers or lost their lives in an attempt to do so. Anything from Athenian Ostracism through the Crusades to the Spanish Conquistadores fit this usually very violent category. An exceptional window of opportunity opened during the age of liberalism of the 19th century, which created the first wave of globalisation with radically decreasing transportation costs, and an increasing need for primary products and staple foods. This development allowed for immense labour migration and large capital inflows into what is today the USA, Canada, Australia, Argentina, Brazil, and New Zealand. For decadal migration from 1880 to 1914, please see Table 2 in \cite{Green1976} or Table 16 in \cite{Baldwin1999}. The incredible increases in the number of people of European origin in these countries very much changed the entire cultural, institutional, and socio-economic make up of those countries in just a few decades. This window of opportunity resolutely shut during the First World War symbolized by the (first) Red Scare in the USA after which the entire developed world closed its borders and allowed in only small numbers of immigrants under ``Skilled Immigration Points Requirements'' or similar systems. Nevertheless, this change of population and culture of the 19th century imprinted on the minds of both sender and receiver countries the idea of large migrations as something benign, even desirable, or worse, a sort of universal right. This is unfortunate as today the situation has changed radically, including the possible motivation for migration \citep{Borjas1999, Boeri2010}, which suggests any large migration will lead to cultural diversity.

As some form of assimilation, albeit very slow, is continuously taking place, in order for cultural diversity to exist, the level of immigration to a country must be over a long period and significant in numbers. Therefore, it is a deliberate policy decision.

Culture is defined by Hofstede as the programming of the human mind by which one group of people distinguishes itself from another group. It is a very persistent shared phenomenon which rarely changes dramatically in the lifetime of a human being \citep{Hofstede2010, Matei2018}. In contrast, the creation of institutions, while a lengthy process, which can be blocked or reversed in adverse conditions, is a much more rapid development \citep{North1992a, North1994, Olson1996}. As such, it is reasonable to hypothesize that the culture of newcomers will change the institutions of the host country rather than the other way around, provided the newcomers have the possibility to do so. This seems to be particularly the case in democratic societies, which do not promote but limit the assimilation process (cultural pluralism, or multiculturalism).

As culture does not change easily, so even migrants do not change their culture irrespective of where they are. This means that the overall culture of the host country will change. Our paper aims to find out how much and what impact this change will have on the environment surrounding people in host countries. We focus on arguably the most important part of the environment, which is the quality of surrounding governance.

The impact of those changes is larger than what the size of the foreign-born population would suggest. Clearly the whole population reacts to newcomers in a way which causes interaction and creates significant changes in the culture of the host country. We cannot, however, specify to what extent the changes are due to the presence of the foreign-born population directly and to what extent the changes in the host country's culture come about due to the majority population's, or rather both cultures', interaction. The nature of the data suggests the reaction by the majority population supersedes the change stemming from the mere presence of the foreign-born population.

\subsection{Impact of Cultural Diversity on Governance}
\label{sec:theoryDiversity}

The debate on cultural diversity is very old, indeed, recorded at least from the time of classical liberalism. John Stuart Mill declared the following in 1862 regarding the need of nationalism and liberalism for democracy: `Free institutions are next to impossible in a country made up of different nationalities. Among a people without fellow-feeling, especially if they read and speak different languages, the united public opinion necessary to the working of representative government cannot exist. The influences which form opinions and decide political acts are different in the different sections of the country' \citep[p.\ 310]{Mill1862}. And further, Mill declares the reasoning why under duress such society will not hold on to their free institutions: ‘Even if all are aggrieved, none feel that they can rely on the others for fidelity in a joint resistance; the strength of none is sufficient to resist alone, and each may reasonably think that it consults its own advantage most by binding for the favour of the government against the rest' \citep[p.\ 311]{Mill1862}. After one and a half centuries, researchers empowered by statistical analysis seem to consider Mill right. He correctly identified that cultural diversity causes political instability and internal conflict (\citealp{Fearon1996, Easterly1997, Easterly2001, Nettle2007}, i.a.), diminishes trust among different groups and the formation of social capital (\citealp{Montgomery1991, Taylor2000, Zak2001, Greve2003, Putnam2007}, i.a.) while increasing the role of government at the same time as it decreases the provision of public goods \citep{Alesina1999, Miguel2004}. As for Mill's concern about free institutions and what we call citizens' rights today, it is safe to say that various diversities make maintaining those institutions and rights significantly more difficult \citep{Krain1998, Walker2002}.

More to the point, some cultures are better at productive enterprises and economic creativity, than others \citep{Williams2010}, or as \citet[p.\ 276]{Landes1998} put it: `institutions and culture first; money next'. Also, as Chakraborty observed \citep{Chakraborty2015}, if prevalent national culture has an anti-capitalist bias it would have a debilitating effect on the economic activity. Mixing cultures does not seem to be good for business either, as there is also proven negative correlation between cultural heterogeneity and economic performance measured by either GDP, growth or productivity (\citealp{Pool1972, Lian1997, Nettle2000, Grafton2002, Alesina2003}, i.a.). Finally, international investment can be significantly affected when actors from several countries meet, as culture seems to define much of these encounters. \cite{Rejchrt2015} suggest that with high power distance in their home countries, non-domestic companies will likely not be compliant with host countries' corporate governance codes. \cite{Barkema1997} claim differences in cultural dimensions do lead to untimely dissolutions of international joint ventures. Governance, and property protection in particular, seems to be a decisive factor for direct or portfolio foreign investment (FDI versus FPI) with the former dominating in countries with low protection and vice versa \citep{Wu2012}.

However, the literature also defines two areas where cultural diversity might be helpful to society. Both cases fit the economic concept of increased consumer choice where even countries producing the same product, say motorcycles, can trade profitably, so that American consumers can enjoy variety enhanced by German brands of motorcycles and vice versa. One of these areas where cultural diversity can be helpful is innovation. It has been accepted that highly individualistic cultures with low power distance as well as low uncertainty avoidance and long-term orientation are better in innovation (\citealp{Gorodnichenko2011, Taylor2012}, i.a.). As there are many other factors that play a part in the process of innovation such as educational levels, levels of infrastructure or spatial concentration and social milieus \citep{Grozinger2017}, it is not easy to separate out the effects of diversity. Moreover, studies on the relationship between innovation and diversity are not common. Different cognitive models and the experience of a diverse pool of researchers can bring about better creative outcomes \citep{Niebuhr2010, Stahl2010} while it can also lead to disagreements and a lack of organization \citep{Ostergaard2011, Harvey2013a}. \cite{Zhan2015} suggest that those opposing results are because of researchers using race, ethnicity, nationality and culture interchangeably in their specifications. They claim that ethnic diversity dampens innovation input thus impairing innovation output but if ethnic diversity is sufficiently low, cultural diversity can increase innovation output. If the results of this paper can be replicated on a larger sample size it might change HR and admission policies for universities and research facilities alike.

The second area where the impact of cultural diversity is not clearly negative and has thus been the subject of hundreds of studies, is the functioning of work groups and labour market workings in general. The ways in which diversity influences the workplace are well described. On one hand, higher diversity increases the effectiveness of tasks that call for a variety of viewpoints and experiences, which has been the main finding of many lab and academic setting type studies \citep{Buller1986, Cox1991, Watson1993}. However, after years of glorification of diversity penned particularly by American researchers advising that managers should increase workforce diversity to enhance work group effectiveness, a more balanced and critical approach has been chosen by the majority of authors. The field studies which were not based on value-in-diversity but rather social-identity perspectives suggested that increased cultural diversity caused a lack of trust and more barriers to social interaction, which in turn instigated low organization and the inhibited productivity of diverse groups \citep{Williams1998, Pelled1999, Earley2000, Richard2004}. Some authors \citep{Ely2001} try to resolve the issue by setting conditions under which cultural diversity would enhance work group functioning and productivity, but the issue remains unresolved with researchers being mostly sceptical about the prospects of some general recommendations.

Based on the economic rule noted above, the fractionalization of some western societies does not have the same impact on every area or group of those societies. \cite{AwaworyiChurchill2017} for example claims that despite the fact that `the very fundamentals of building of good institutions are impaired by diversity' (p.\ 593) and ‘ethnic and linguistic fractionalization negatively influences entrepreneurship (new business density), and the institutional environment affecting entrepreneurship' (p.\ 590), there might be groups to benefit; as they might be accustomed to it. These are ethnic groups where business ownership and self-employment are more pronounced as some entrepreneurs and investors choose to do business in highly fractionalized areas \cite{Khayesi2014}. Also, \cite{Ottaviano2006} suggests there is a silver lining to the increased levels of diversity in the United States. While immigration caused lower wages particularly for the low-skilled workforce, labour market fractionalization and possibly endangered cultural values for the ``natives'', US metropolitan areas with an increased cultural diversity experienced higher wages and rental prices of their housing, as well as an increased variety of consumption.

\section{Measures of Governance and Culture}
\label{sec:measure}

\subsection{Worldwide Governance Indicators}
\label{sec:measureGovernance}

Governance, or the way authority in a country is exercised, is among the most important institutions for one's well-being. The way how governments are selected and monitored and how efficiently and transparently they formulate and implement their policies is crucial for much of the country's development.

The Worldwide Governance Indicators (WGI) were developed by the World Bank and are intended to provide a comprehensive assessment of the quality of governance in different countries around the world. The indicators are measured through a composite index that combines data from various sources related to governance, including surveys of experts, assessments by international organizations, and other objective measures of governance performance (see \citealp{Kaufmann2011}). Each indicator is standardized to have a mean of zero and a standard deviation of one, resulting in a range of approximately -2.5 to 2.5. The higher the score, the better the governance performance of a country relative to others. The six governance indicators are defined in the following way:
\begin{itemize}
\item \emph{Voice and Accountability (VA)} captures perceptions of the extent to which a country's citizens are able to participate in selecting their government, as well as freedom of expression, freedom of association, and a free media.  
\item \emph{Political Stability and Absence of Violence/Terrorism (PV)} captures perceptions of the likelihood that the government will be destabilized or overthrown by unconstitutional or violent means, including politically‐motivated violence and terrorism.
\item \emph{Government Effectiveness (GE)} captures perceptions of the quality of public services, the quality of the civil service and the degree of its independence from political pressures, the quality of policy formulation and implementation, and the credibility of the government's commitment to such policies.
\item \emph{Regulatory Quality (RQ)} captures perceptions of the ability of the government to formulate and implement sound policies and regulations that permit and promote private sector development.
\item \emph{Rule of Law (RL)} captures perceptions of the extent to which agents have confidence in and abide by the rules of society, and in particular the quality of contract enforcement, property rights, the police, and the courts, as well as the likelihood of crime and violence.
\item \emph{Control of Corruption (CC)} captures perceptions of the extent to which public power is exercised for private gain, including both petty and grand forms of corruption, as well as “capture” of the state by elites and private interests.
\end{itemize}

\begin{figure}
\begin{center}
\includegraphics[height=8.3cm]{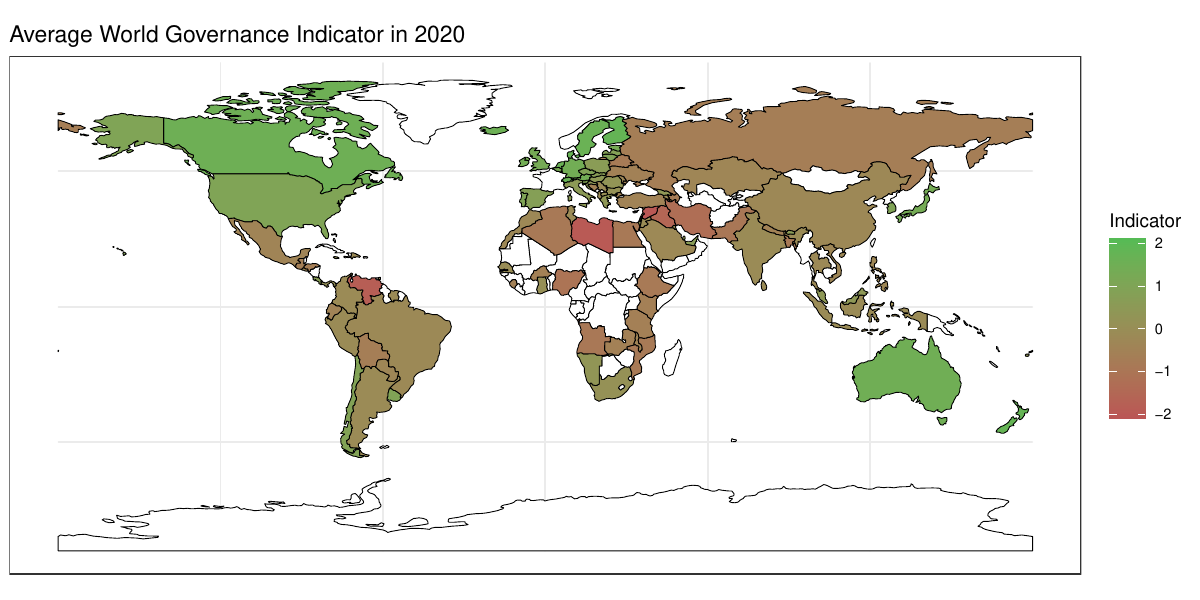}
\caption{The average world governance indicator in the analysed countries in 2020.}
\label{fig:worldWgi}
\end{center}
\end{figure}

There is a strong correlation among the six governance indicators suggesting they measure the same broad concept \citep{Langbein2010}. Consequently, some studies average the six indicators into a single index \citep{Al-Marhubi2004, Bjornskov2006}. Figure \ref{fig:worldWgi} shows the average Worldwide Governance Indicators in 2020. In our study, we distinguish between the six indicators but analyse them within a single model which takes into account a correlation between them.

\subsection{Hofstede's Cultural Dimensions}
\label{sec:measureHofstede}

Culture has a profound impact on a great variety of institutions and realities around us starting from diverse political and socio-economic development to particularities such as homicide levels \citep{LappiSeppala2015}, economic systems \citep{Pryor2007}, tax structures \citep{Koenig2012}, proclivity to insurance coverage \citep{Outreville2018}, or level of terrorist activity \citep{Meierrieks2013}; please see \cite{Evan2021, Holy2022d} for more references. Cultural dimensions as defined by \cite{Hofstede2010} have an important advantage of being a rather general phenomenon, espousing quite a lot of salient divisions, and is thus a meaningful tool to capture the variety of socio-economic differentiation as its popularity attests.

Hofstede's cultural dimensions are measured using surveys that ask individuals to rate their agreement or disagreement with various statements or questions related to cultural values. The surveys are administered to a sample population from a particular country, and the responses are then analyzed to identify the dominant cultural values within that society. Each dimension is scored on a scale ranging from 0 to 100, with higher scores indicating a stronger emphasis on that particular cultural value. In our study, we rescale the dimensions to a (0, 1) range in order to have a similar magnitude as the Worldwide Governance Indicators; however, this does not change the relative positions of different cultures along each dimension. \cite{Hofstede2010} define the six cultural dimensions in the following way:
\begin{itemize}
\item \emph{Power Distance (PDI)} is the extent to which the less powerful members of institutions and organizations within a country expect and accept that power is distributed unequally.
\item \emph{Individualism (IDV)} pertains to societies in which the ties between individuals are loose: everyone is expected to look after him- or herself and his or her immediate family. Collectivism as its opposite pertains to societies in which people from birth onward are integrated into strong, cohesive in-groups, which throughout people's lifetime continue to protect them in exchange for unquestioning loyalty.
\item \emph{Masculinity (MAS)} refers to society where emotional gender roles are clearly distinct: men are supposed to be assertive, tough, and focused on material success, whereas women are supposed to be more modest, tender, and concerned with the quality of life. A society is called feminine when emotional gender roles overlap: both men and women are supposed to be modest, tender, and concerned with the quality of life.
\item \emph{Uncertainty Avoidance (UAI)} is the extent to which the members of a culture feel threatened by ambiguous or unknown situations.
\item \emph{Long-Term Orientation (LTO)} stands for the fostering of virtues oriented toward future rewards -- in particular, perseverance and thrift. Its opposite pole, short-term orientation, stands for the fostering of virtues related to the past and present -- in particular, respect for tradition, preservation of ``face'', and fulfilling social obligations.
\item \emph{Indulgence (IVR)} stands for a tendency to allow relatively free gratification of basic and natural human desires related to enjoying life and having fun. Its opposite pole, restraint, reflects a conviction that such gratification needs to be curbed and regulated by strict social norms.
\end{itemize}
    
Hofstede's cultural dimensions have attracted criticism (\citealp{Dorfman1988, Schwartz1999, McSweeney2002, Jones2007, Taras2016}, i.a.) as among other things they suppose that within a state, there is a uniform national culture. This is rather paradoxical as Hofstede himself started his research about cultural differences not only within one country but within one corporation. Many of his well-known quotes are painfully aware of the dangers of the close proximity of different cultures: `Culture is more often a source of conflict than of synergy. Cultural differences are a nuisance at best and often a disaster'\footnote{As quoted by \citet[p.\ 37]{Vance2015}.}. This nuisance or disaster is clearly present within countries, as few of them are completely culturally uniform. There are several ways how to deal with the problem. \cite{House2004} and \cite{Taras2016} suggest doing away with countries, as their borders are not good boundaries of culture, in favour of professions or socio-economic classes. Others think the boundaries of states are useful when measuring culture: `Moreover, nations matter because they have governments who establish policies on trade, factor mobility, and the like (which policies, it can be argued, are shaped by the national culture)' \citep[p.\ 408]{Williams2010}.

There are several alternative survey-based measures of culture available in the literature, including the Schwartz's Basic Human Values (see, e.g., \citealp{Schwartz2012}), the Global Leadership and Organizational Behavior Effectiveness (GLOBE) framework (see e.g., \citealp{House2004}), and the World Values Survey (see, e.g., \citealp{Inglehart2004}). While these measures provide valuable insights into cultural differences, the Hofstede's cultural dimensions framework is still the dominant approach in the field, as evidenced by literature reviews of \cite{Kirkman2006} and \cite{Beugelsdijk2017}. One compelling reason to use Hofstede's cultural dimensions is the availability of data for a large number of countries, which is provided by \cite{Hofstede2021}. This comprehensive dataset allows for cross-cultural comparisons and analysis on a global scale, making it a valuable resource for researchers and practitioners alike.

\subsection{Cultural Diversity Measures}
\label{sec:measureDiversity}

The existing body of literature on cultural diversity is also large but it is less compact, given the variety of different approaches. Firstly, different diversities were used in terms of (i) race and ethnicity, (ii) religion, (iii) language, (iv) or their combinations, and (v) culture. In order to distinguish the impact of culture on institutions using secondary data, researchers used comparison of the diversities named above largely with major socio-economic variables. However, geographic variables \citep{Laitin2012}, changes in institutions stemming from changed migration patterns or levels of assimilation \citep{Algan2010, Luttmer2011}, or new epidemiological approaches \citep{Fernandez2010, Maseland2013} were also used.

There are two measurements of cultural diversity used most often. Both have attracted significant criticism. The cultural polarisation concept has been criticised for being vague as \cite{Bramson2017} summarizes: `A range of very different social configurations and very different social dynamics have been lumped together under the term ``polarization''.' (p.\ 117). And the Ethno-Linguistic Fractionalisation index, or ELF -- the likelihood that two people chosen at random will be from different ethnic groups -- has been criticized for having at least four major problems \citep{Laitin2001}. While we accept and cite from findings based on those methods in the theoretical part, we have chosen a different approach in our empirical study. We define our cultural diversity measure based on Hofstede's cultural dimensions and the international migrant stock with the tools of basic statistical analysis.

\section{Proposed Model}
\label{sec:model}

\subsection{Cultural Level and Diversity Indicators}
\label{sec:modelIndicator}

A disadvantage of Hofstede's cultural dimensions theory is that it assumes that the culture is homogeneous in each country. We relax this assumption and let the culture be heterogenous with respect to the different countries of origin of the citizens. We define two measures to capture the heterogeneity of the culture.

First, we assign each citizen of a country Hofstede's cultural dimensions of the country he or she was born in. Let $POP_{i,t}$ denote the population of country $i$ as of year $t$ and $BIC_{i,t,o}$ the number of citizens in country $i$ born in country $o$ as of year $t$. Furthermore, let $HCD_{i,k}$ denote the $k$-th Hofstede's cultural dimension of country $i$. The $k$-th cultural level indicator $CLI_{i,t,k}$ of country $i$ as of year $t$ is defined as the average of the $k$-th Hofstede's cultural dimension over all citizens of the country, i.e.
\begin{equation}
CLI_{i,t,k} = \sum_{o=1}^N \frac{BIC_{i,t,o}}{POP_{i,t}} HCD_{o,k}.
\end{equation}
Note that Hofstede's cultural dimensions are constant over time while our proposed indicators are dynamic due to the changing structure of citizens.

\begin{figure}
\begin{center}
\includegraphics[height=8.3cm]{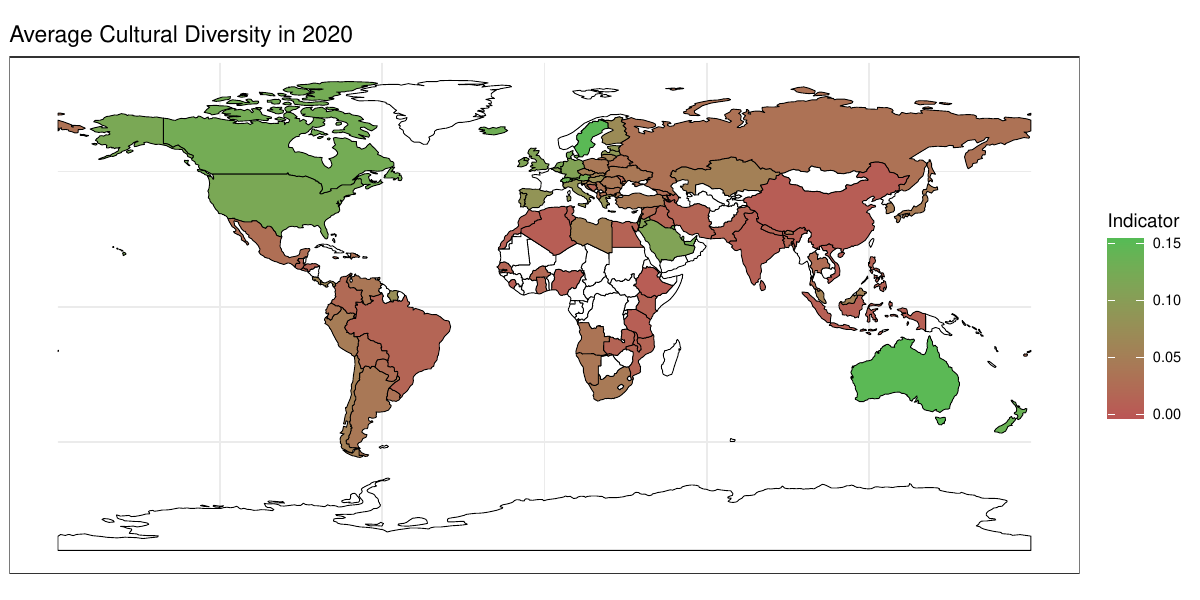}
\caption{The average cultural diversity measure in the analysed countries in 2020.}
\label{fig:worldDiversity}
\end{center}
\end{figure}

The $k$-th cultural diversity indicator $CDI_{i,t,k}$ of country $i$ as of year $t$ is defined as the standard deviation of the $k$-th Hofstede's cultural dimension over all citizens of the country, i.e.
\begin{equation}
CDI_{i,t,k} = \sqrt{\sum_{o=1}^N \frac{BIC_{i,t,o}}{POP_{i,t}} \left(HCD_{o,k} - CLI_{i,t,k} \right)^2 }.
\end{equation}
Values close to zero indicate that all citizens have similar Hofstede's cultural dimensions (given by the cultural level indicator) while larger values indicate that Hofstede's cultural dimensions of citizens are dispersed (some are lower and some higher than the cultural level indicator). Figure \ref{fig:worldDiversity} shows the cultural diversity indicator averaged over the six dimensions in 2020.

Naturally, there are limitations mainly due to a lack of data, allowing us only to study migration patterns over the last few decades. An admixture of earlier migrations or even centuries-old minorities in the countries of our sample are taken simply as part of the original culture of the host country. Thus, the heterogeneity relates only to recent migrations. We are also unable to identify holders of what culture migrated from one country to another, due to a lack of data. In another words, if a French citizen of German ethnicity from Alsace managed to hold to German culture and has migrated to the Netherlands, our model will nevertheless consider him a Frenchman in the Netherlands.

\subsection{Regression Model of Governance}
\label{sec:modelRegression}

Our aim is to model dependence of the Worldwide Governance Indicators on the proposed cultural level and diversity indicators. We let each governance indicator have its own equation, but we use a single model as we allow for correlations between the governance indicators.

Let $WGI_{i,t,j}$ denote the $j$-th governance indicator of country $i$ as of year $t$. The regression equations are then given by
\begin{equation}
WGI_{i,t,j} = \alpha_j + \sum_{k=1}^K \beta_{j,k} CLI_{i,t,k} + \sum_{k=1}^K \gamma_{j,k} CDI_{i,t,k} + \sum_{l=1}^L \delta_{j,l} Z_{i,t,l} + u_{i,t,j},
\end{equation}
where $\alpha_i$ are intercepts, $\beta_{j,k}$ are coefficients for the cultural level indicators $CLI_{i,t,k}$, $\gamma_{j,k}$ are coefficients for the cultural diversity indicators $CDI_{i,t,k}$, $\delta_{j,l}$ are coefficients for control variables $Z_{i,t,l}$, and $u_{i,t,j}$ is the error term. We let the error term have quite a rich dependence structure allowing for correlations in all three dimensions. First, we assume a spatial dependence in the form of a spatial error model (SEM) with the weight matrix given by the portions of immigrants from the individual countries (see, e.g., \citealp[Section 11.7]{Greene2018}). Second, we assume a temporal dependence in the form of an autoregressive process (AR) of the first order (see, e.g., \citealp[Section 20.3]{Greene2018}). Third, we assume a dependence between the equations in the form of seemingly unrelated regressions (SUR) (see, e.g., \citealp[Section 10.2]{Greene2018}). All put together, the error term has the structure 
\begin{equation}
u_{i,t,j} = \lambda_j \sum_{o \neq i} \frac{BIC_{i,t,o}}{POP_{i,t} - BIC_{i,t,i}} u_{o,t,j} + \varphi_j u_{i,t-1,j} + e_{i,t,j},
\end{equation}
where $\lambda_j$ are spatial coefficients, $\varphi_j$ are serial coefficients, and $e_{i,t,j}$ is a normally distributed random variable with zero mean and covariance given by
\begin{equation}
\mathrm{E} \left[ e_{i_1,t_1,j_1} e_{i_2,t_2,j_2} \right] = \begin{cases}
0 & \text{for } i_1 \neq i_2 \text{ or } t_1 \neq t_2, \\
\sigma_{j_1,j_2} & \text{for } i_1 = i_2 \text{ and } t_1 = t_2. \\
\end{cases}
\end{equation}

We estimate the model by the maximum likelihood method. Note that modelling the governance indicators together in the SUR fashion increases the efficiency of the estimator in contrast to modelling each equation separately.

\section{Empirical Study}
\label{sec:est}

\subsection{Data Sample}
\label{sec:estData}

As noted above, the construction of the proposed cultural level and diversity indicators is limited due to data availability. Our source of Hofstede's six cultural dimensions is \cite{Hofstede2021}, which offers data for 118 countries. The data on international migrant stock are obtained from the \cite{InternationalMigrantStock2021} and are available for 232 countries and regions in 1990, 1995, 2000, 2005, 2010, 2015, and 2020\footnote{The \cite{InternationalMigrantStock2021} source of our data uses the following explanatory note (p.\ 3): `International migrants have been equated with the foreign-born population whenever this information is available, which is the case in most countries or areas. In most countries lacking data on place of birth, information on the country of citizenship of those enumerated was available and was used as the basis for the identification of international migrants, thus effectively equating, in these cases, international migrants with foreign citizens.'}. The data also include some immigrants with unspecified countries of origin.

To construct the proposed cultural level and diversity indicators, we would need to know Hofstede's cultural dimensions of all the countries of origin of all immigrants. Unfortunately, not all are available. For the purposes of our indicators, we resort to the spatial imputations of Hofstede's cultural dimensions. Each unknown value is interpolated as the average of the five nearest neighbours\footnote{Other methods, such as those using weights based on inverse distance, inverse square distance, or different numbers of neighbours, lead to very similar results suggesting the robustness of our approach.}. We stress that we use this approach only for the construction of the proposed indicators with foreign countries of origin. When a country has unknown Hofstede's cultural dimensions, we do not compute the proposed indicators for it and exclude it from the regression model. Concerning the portion of immigrants with an unknown country of origin, we assign them individual countries proportionally to the total number of emigrants from the individual countries. These two data imputation steps are not ideal, but are necessary.

The source of the Worldwide Governance Indicators is the \cite{WGI2021}. The governance indicators are compiled biannually from 1996 to 2002 and annually since 2002. They are available for 214 countries and territories, with some missing values.

As control variables, we consider the three constituents of the Human Development Index (HDI): the life expectancy at birth (LE), the education index combining the mean years of schooling completed and the expected years of schooling upon entering the education system (SY), and the gross national income per capita measured in purchasing power standards (GNI). To ensure comparability across years, we use the ranks of each variable rescaled to a range of (0, 1) for each year. The source of the data is \cite{HumanDevelopmentIndex2021}.

As we merge data from various sources with different samples of countries and time frames, some observations are lost. Our final data sample consists of $K=6$ cultural dimensions, $M=6$ Worldwide Governance Indicators, and $L=3$ control variables, observed for $N=115$ countries in $T=5$ time periods covering the years 2000, 2005, 2010, 2015, and 2020. The list of countries together with summarized values values of Worldwide Governance Indicators, Hofstede’s cultural levels, and Hofstede’s cultural diversities are presented in Table \ref{tab:country} in the appendix.

\subsection{Model Specification}
\label{sec:estSpec}

As the first step, we check whether the assumed error structure in our model is reasonable. Table \ref{tab:aic} reports the Akaike information criterion (AIC) of the model with uncorrelated errors, the model with only spatially correlated errors, the model with only serially correlated errors, the model with only SUR errors, and the model described in Section \ref{sec:modelRegression} with all three error structures. According to Table \ref{tab:coef}, the spatial parameter $\lambda_j$ is positive for all six governance indicators and significant for five. Table \ref{tab:aic} shows a small decrease of AIC when assuming spatially correlated errors rather than uncorrelated errors. Spatial dependence is therefore present in the error term but appears to be rather weak. Serial correlation, on the other hand, is quite strong. According to Table \ref{tab:coef}, the serial parameter $\varphi_j$ ranges between 0.72 and 0.87 for the individual governance indicators and is significant in all cases. The decrease of AIC in Table \ref{tab:aic} is quite substantial when assuming serially correlated errors rather than uncorrelated errors. A similarly high decrease is observed when assuming correlation between the individual indicators rather than uncorrelated errors. According to Table \ref{tab:corr}, the correlations of inter-equation residuals range between 0.29 and 0.65. We therefore proceed with the model that assumes all three error structures, as described in Section \ref{sec:modelRegression}.

\begin{table}
\begin{center}
\caption{The Akaike information criterion for various explanatory variables (in rows) and error structures (in columns).}
\label{tab:aic} 
\scriptsize
\begin{tabular}{lccccc} 
\toprule
& Indep. & Spatial & Serial & SUR & All \\
\midrule 
Hofstede's Cultural Dimensions    & \num{4988.76} & \num{4891.77} & \num{1826.69} & \num{2677.70} &  \num{500.57} \\ 
Heterogeneous Level               & \num{4997.16} & \num{4925.11} & \num{1831.97} & \num{2652.21} &  \num{495.90} \\ 
Heterogeneous Level and Diversity & \num{4887.11} & \num{4833.97} & \num{1796.69} & \num{2544.97} &  \num{481.67} \\ 
\bottomrule
\end{tabular}
\end{center}
\end{table}

Next, we consider Hofstede's cultural dimensions to be the explanatory variables in our model, similarly to our previous work in \cite{Holy2022d}\footnote{Note that the model of \cite{Holy2022d} is different as it contains an additional error term representing inefficiency of the individual countries in the sense of the stochastic frontier analysis and, on the other hand, assumes that the errors are uncorrelated.}. We compare the fit of this base model with the model in which Hofstede's cultural dimensions are replaced by the proposed cultural level indicators. According to Table \ref{tab:aic}, the use of the proposed heterogeneous indicators leads to slightly better fit as its AIC is lower by 4.68 in the case of the model with the richest error structure. Both models, however, perform almost identically in terms of the coefficient of determination, as they both have adjusted R$^2$ of 0.716. When the original dimensions are replaced by the cultural level indicators and the cultural diversity indicators are further added, the AIC improves by 18.90. The adjusted R$^2$ of this model is 0.723, which is only slightly higher. Nonetheless, the proposed model outperforms the alternatives in both of the considered metrics. In the sequel, we work only with the proposed model containing both the cultural level and diversity indicators.

\subsection{Relation Between Governance and Culture}
\label{sec:estRelation}

\begin{figure}
\begin{center}
\includegraphics[height=8.3cm]{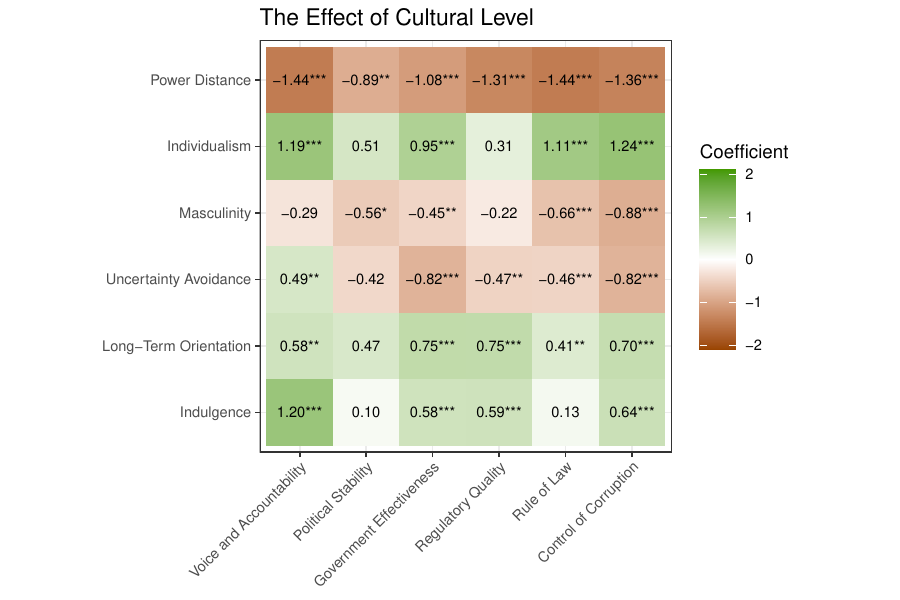}
\caption{The estimated coefficients for cultural level variables in the model with heterogeneous level and diversity. The color indicates the values of the coefficients.}
\label{fig:coefLevel}
\end{center}
\end{figure}

\begin{figure}
\begin{center}
\includegraphics[height=8.3cm]{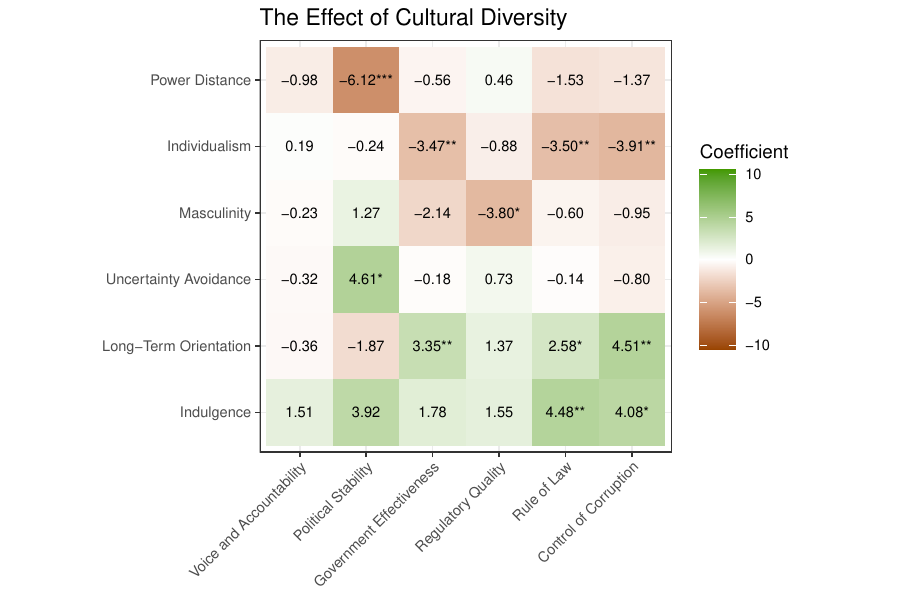}
\caption{The estimated coefficients for cultural diversity variables in the model with heterogeneous level and diversity.}
\label{fig:coefDiversity}
\end{center}
\end{figure}

The estimated coefficients of the final model are presented in Table \ref{tab:coef}, and, for better readability, also in Figures \ref{fig:coefLevel} and \ref{fig:coefDiversity}. Note that the coefficients for the cultural level indicators are directly comparable among themselves as these indicators are standardized on the same scale. Likewise, the coefficients for the cultural diversity indicators are also comparable amongst themselves. The coefficients for level indicators, however, are not comparable to the coefficients for diversity indicators. The color coding in our Figures \ref{fig:coefLevel} and \ref{fig:coefDiversity} represents the values of the coefficients. We summarize our findings in Table \ref{tab:summary}.

\begin{table}
\begin{center}
\caption{The estimated coefficients and their standard errors (in parentheses) in the model with heterogeneous level and diversity.}
\label{tab:coef} 
\scriptsize
\begin{tabular}{lcccccc} 
\toprule
& VA & PV & GE & RQ & RL & CC \\
\midrule
Constant & $-$0.75$^{**}$ & $-$0.08 & $-$0.18 & $-$0.11 & 0.18 & 0.12 \\ 
  & (0.26) & (0.34) & (0.21) & (0.25) & (0.21) & (0.23) \\ 
  & & & & & & \\ 
PDI Level $\beta_{j,1}$ & $-$1.44$^{***}$ & $-$0.89$^{**}$ & $-$1.08$^{***}$ & $-$1.31$^{***}$ & $-$1.44$^{***}$ & $-$1.36$^{***}$ \\ 
  & (0.24) & (0.31) & (0.19) & (0.22) & (0.19) & (0.21) \\ 
  & & & & & & \\ 
IDV Level $\beta_{j,2}$ & 1.19$^{***}$ & 0.51 & 0.95$^{***}$ & 0.31 & 1.11$^{***}$ & 1.24$^{***}$ \\ 
  & (0.24) & (0.31) & (0.19) & (0.22) & (0.19) & (0.21) \\ 
  & & & & & & \\ 
MAS Level $\beta_{j,3}$ & $-$0.29 & $-$0.56$^{*}$ & $-$0.45$^{**}$ & $-$0.22 & $-$0.66$^{***}$ & $-$0.88$^{***}$ \\ 
  & (0.18) & (0.24) & (0.14) & (0.17) & (0.15) & (0.16) \\ 
  & & & & & & \\ 
UAI Level $\beta_{j,4}$ & 0.49$^{**}$ & $-$0.42 & $-$0.82$^{***}$ & $-$0.47$^{**}$ & $-$0.46$^{***}$ & $-$0.82$^{***}$ \\ 
  & (0.16) & (0.21) & (0.13) & (0.15) & (0.13) & (0.14) \\ 
  & & & & & & \\ 
LTO Level $\beta_{j,5}$ & 0.58$^{**}$ & 0.47 & 0.75$^{***}$ & 0.75$^{***}$ & 0.41$^{**}$ & 0.70$^{***}$ \\ 
  & (0.18) & (0.24) & (0.15) & (0.17) & (0.15) & (0.16) \\ 
  & & & & & & \\ 
IND Level $\beta_{j,6}$ & 1.20$^{***}$ & 0.10 & 0.58$^{***}$ & 0.59$^{***}$ & 0.13 & 0.64$^{***}$ \\ 
  & (0.18) & (0.23) & (0.14) & (0.16) & (0.14) & (0.16) \\ 
  & & & & & & \\ 
PDI Diversity $\gamma_{j,1}$ & $-$0.98 & $-$6.12$^{***}$ & $-$0.56 & 0.46 & $-$1.53 & $-$1.37 \\ 
  & (1.15) & (1.48) & (0.91) & (1.07) & (0.93) & (1.02) \\ 
  & & & & & & \\ 
IDV Diversity $\gamma_{j,2}$ & 0.19 & $-$0.24 & $-$3.47$^{**}$ & $-$0.88 & $-$3.50$^{**}$ & $-$3.91$^{**}$ \\ 
  & (1.58) & (2.03) & (1.24) & (1.46) & (1.27) & (1.39) \\ 
  & & & & & & \\ 
MAS Diversity $\gamma_{j,3}$ & $-$0.23 & 1.27 & $-$2.14 & $-$3.80$^{*}$ & $-$0.60 & $-$0.95 \\ 
  & (1.69) & (2.13) & (1.30) & (1.54) & (1.34) & (1.46) \\ 
  & & & & & & \\ 
UAI Diversity $\gamma_{j,4}$ & $-$0.32 & 4.61$^{*}$ & $-$0.18 & 0.73 & $-$0.14 & $-$0.80 \\ 
  & (1.50) & (1.91) & (1.17) & (1.38) & (1.20) & (1.31) \\ 
  & & & & & & \\ 
LTO Diversity $\gamma_{j,5}$ & $-$0.36 & $-$1.87 & 3.35$^{**}$ & 1.37 & 2.58$^{*}$ & 4.51$^{**}$ \\ 
  & (1.59) & (2.08) & (1.28) & (1.50) & (1.29) & (1.42) \\ 
  & & & & & & \\ 
IND Diversity $\gamma_{j,6}$ & 1.51 & 3.92 & 1.78 & 1.55 & 4.48$^{**}$ & 4.08$^{*}$ \\ 
  & (1.85) & (2.43) & (1.51) & (1.76) & (1.51) & (1.66) \\ 
  & & & & & & \\ 
LE Control $\delta_{j,1}$ & 0.53$^{**}$ & 0.40 & 0.61$^{***}$ & 0.64$^{***}$ & 0.74$^{***}$ & 0.66$^{***}$ \\ 
  & (0.17) & (0.23) & (0.14) & (0.16) & (0.14) & (0.15) \\ 
  & & & & & & \\ 
SY Control $\delta_{j,2}$ & 0.44$^{**}$ & 0.07 & 0.21 & 0.22 & 0.01 & 0.09 \\ 
  & (0.16) & (0.22) & (0.14) & (0.16) & (0.14) & (0.15) \\ 
  & & & & & & \\ 
GNI Control $\delta_{j,3}$ & $-$0.03 & 1.30$^{***}$ & 1.22$^{***}$ & 1.02$^{***}$ & 1.00$^{***}$ & 0.95$^{***}$ \\ 
  & (0.21) & (0.28) & (0.17) & (0.20) & (0.17) & (0.19) \\ 
  & & & & & & \\ 
Spatial Parameter $\lambda_j$ & 0.06 & 0.16 & 0.09 & 0.09 & 0.12 & 0.10 \\ 
  & (0.03) & (0.03) & (0.04) & (0.04) & (0.03) & (0.04) \\
  & & & & & & \\ 
Serial Parameter $\varphi_j$  & 0.87 & 0.76 & 0.72 & 0.75 & 0.79 & 0.77 \\ 
  & (0.02) & (0.03) & (0.03) & (0.03) & (0.02) & (0.03) \\
  & & & & & & \\ 
\bottomrule
\multicolumn{7}{r}{\textit{Note:} $^{***}p < 0.001$; $^{**}p < 0.01$; $^{*}p < 0.05$}
\end{tabular}
\end{center}
\end{table}

\begin{table}
\begin{center}
\caption{The variance (diagonal), covariance (lower triangle), and correlation (upper triangle) of inter-equation residuals in the model with heterogeneous level and diversity.}
\label{tab:corr} 
\scriptsize
\begin{tabular}{lcccccc} 
\toprule
& VA & PV & GE & RQ & RL & CC \\
\midrule
VA & \GC{0.10} & 0.29 & 0.40 & 0.45 & 0.53 & 0.43 \\ 
PV & \GC{0.04} & \GC{0.18} & 0.35 & 0.29 & 0.44 & 0.36 \\ 
GE & \GC{0.03} & \GC{0.04} & \GC{0.07} & 0.63 & 0.65 & 0.59 \\ 
RQ & \GC{0.04} & \GC{0.04} & \GC{0.05} & \GC{0.09} & 0.58 & 0.46 \\ 
RL & \GC{0.04} & \GC{0.05} & \GC{0.05} & \GC{0.05} & \GC{0.07} & 0.64 \\ 
CC & \GC{0.04} & \GC{0.04} & \GC{0.05} & \GC{0.04} & \GC{0.05} & \GC{0.08} \\ 
\bottomrule
\end{tabular}
\end{center}
\end{table}

\begin{table}
\begin{center}
\caption{Summary of the effects in the model with heterogeneous level and diversity.}
\label{tab:summary} 
\scriptsize
\begin{tabular}{lll} 
\toprule
& Effect of Level      & Effect of Diversity  \\
\midrule
Power Distance        & Entirely Negative    & Partly Negative      \\
Individualism         & Mostly Positive      & Partly Negative      \\
Masculinity           & Mostly Negative      & Partly Negative      \\
Uncertainty Avoidance & Mixed                & Partly Positive      \\
Long-Term Orientation & Mostly Positive      & Partly Positive      \\
Indulgence            & Mostly Positive      & Partly Positive      \\
\bottomrule
\end{tabular}
\end{center}
\end{table}

The model includes control variables to account for the level of development in terms of healthcare, education, and economic well-being. We found that life expectancy has a significant positive effect on all governance indicators except Political Stability. The variable measuring school years has mostly insignificant effects, except for a positive effect on Voice and Accountability. The gross national income per capita has a significant positive effect on all governance indicators except Voice and Accountability.

Regarding the cultural level indicators, Power Distance has the most pronounced and negative impact on the quality of institutions across all governance variables. Masculinity also exhibits a negative effect, albeit to a lesser degree than Power Distance and statistically significant only for 4 out of 6 dependent variables. Interestingly, we observed that Uncertainty Avoidance has a significantly positive impact on Voice and Accountability, but a negative effect on other governance indicators. Individualism, Long-Term Orientation, and Indulgence all demonstrate positive effects on all governance indicators, and in most cases, these effects are also statistically significant.

When comparing the previous model of \cite{Holy2022d} with our newly proposed model allowing for admixing of the foreign born population and a more general correlation structure, we can say that the direction of the effects remains the same but the values and significance slightly differs. We believe the proposed model to be more reliable due to the better handling of correlations in the error term. The proposed model also has a better fit as it takes into account the effects of cultural heterogeneity. Typical migration patterns involve people moving from less developed countries to more developed ones. The less developed countries tend to have governance of considerably inferior quality connected to particular cultural roadblocks as identified in \cite{Holy2022d} and confirmed here. Namely, the positive effect of Individualism, Long-Term Orientation, and Indulgence with the negative effect of Power Distance, and to a lesser extent Masculinity and Uncertainty Avoidance. Therefore, the admixture of holders of cultures from less developed countries should cause the deterioration of governance even in developed countries across the board. By incorporating information on migration patterns, cultural level indicators can provide a more accurate measure of culture compared to Hofstede's cultural dimensions.

Heterogeneity of culture is also captured by the cultural diversity indicators. Overall, the effects of diversity are not as pronounced as in the case of the cultural level indicators. Nevertheless, there are some interesting results. Power Distance, Individualism, and Masculinity have a partly significant negative effect on governance and a partly insignificant effect. Most prominently, a large variation in Power Distance among the citizens leads to low Political Stability. Similarly, but to a lesser extent, a large variation in Individualism leads to low Government Effectiveness, Rule of Law, and Control of Corruption, while a large variation in Masculinity leads to low Regulatory Quality. The positive effects of diversity, on the other hand, can be found for Uncertainty Avoidance, Long-Term Orientation, and Indulgence. Specifically, a large variation in Uncertainty Avoidance among citizens leads to high Political Stability, a large variation in Long-Term Orientation leads to high Government Effectiveness, Rule of Law, and Control of Corruption, and a large variation in Indulgence leads to high Rule of Law and Control of Corruption.

Finally, let us examine the results of our analysis in terms of specific governance indicators. Voice and Accountability is primarily affected negatively by Power Distance and positively by Individualism and Indulgence, with a lesser positive effect from Uncertainty Avoidance and Long-Term Orientation. As the only governance indicator, it is not significantly affected by any of our diversity measures. Political Stability is only moderately affected by cultural levels, with negative effects from Power Distance and Masculinity. However, diversity in Power Distance has a very high negative effect, the highest among all diversity coefficients, while variation in Uncertainty Avoidance has a positive effect. Government Effectiveness, Rule of Law, and Control of Corruption exhibit very similar patterns. These are high negative effects of Power Distance, negative moderate effects of Masculinity and Uncertainty Avoidance, high positive effects of Individualism, and moderate positive effects of Long-Term Orientation and Indulgence. Regarding cultural diversity, variation in Individualism has moderate negative effects, while variation in Long-Term Orientation and Indulgence have moderate positive effects. Regulatory Quality is negatively affected by Power Distance and Uncertainty Avoidance and positively affected by Long-Term Orientation and Indulgence. It is the only governance indicator that is affected by variation in Masculinity, with a moderate negative relation.

\subsection{Implications}
\label{sec:estImplications}

The results of our study show that there is a strong and statistically significant negative relationship between Hofstede's cultural dimension of Power Distance and all governance indicators. Put simply, governance cannot be expected to improve if less powerful members of institutions or society as a whole are not convinced of the meaningfulness of any participation in the electoral process or civil society, to give just two examples. To address this issue, targeted programs should be implemented to encourage those disfranchised that their voice and participation are valuable and will be heard. Another unique finding is that cultural diversity in Power Distance has a negative impact on Political Stability. Assimilation - preferably towards a reduction in Power Distance - would increase a country's Political Stability. Newcomers from countries with a high Power Distance cultural indicator should be able to understand that they do not have to endorse inequality or status symbols in society, to name a few examples. Yet this may be difficult to achieve because of persistent cultural practices. Similarly, we found that large differences in Individualism within a population are problematic for good governance. As we also found Individualism to be strongly correlated with good governance as such, it might be a good policy to try to decrease extremely collectivist attitudes, if they exist, by reducing the degree of interdependence among members of the society. On the other hand, cultural diversity on the long-short term orientation scale as well as an indulgence-restrain scale have a statistically significant positive effect on governance and should be promoted.

\cite{Tang2008} argue that dramatic changes in economic conditions can be the source of cultural dynamics even within tens of years. They have found that some of Hofstede's cultural dimensions are less permanent than others. These are Individualism, Long-Term Orientation, and Power Distance, while Uncertainty Avoidance and Masculinity are more stable as they seem to represent traditional institutions such as language, religion, or surrounding conditions. Their results were confirmed over a longer timeframe by \cite{Zhao2016} while the connection of cultural change to (socio-) economic development has also been verified by \cite{Matei2018}. \cite{Beugelsdijk2015} have found that the overall level of Individualism and Indulgence increased while Power Distance decreased. The relative positions of countries, however, remained quite stable. These findings can be relevant for both our results of cultural influence and cultural diversity. Notably, the negative effects of cultural diversity are mainly in ``changeable'' Power Distance and Individualism. This means that given favourable economic conditions and significant time, the impact of those two cultural roadblocks to good governance should ease.

\section{Discussion}
\label{sec:disc}

\subsection{Limitations of the Study}
\label{sec:discLimit}

We work with the assumption that the national culture is stable in time as well as geographically as supported by research cited. However, the existence of other culture holders in countries prior to the time of our migration data or the existence of minorities that are more inclined to migrate than the respective national culture is a clear limitation of our study and requires further research if adequate data could be found for this purpose.

\subsection{Directions for Future Research}
\label{sec:discFuture}

A promising avenue for future research would be to investigate potential contingent effects between the Hofstede's cultural dimensions and their corresponding diversity measures. Furthermore, it could be valuable to investigate potential nonlinear effects between the Hofstede's cultural dimensions and the Worldwide Governance Indicators. Examining these effects could help to deepen our understanding of the complex relationship between culture, diversity, and governance.

While our empirical study focuses on diversity in terms of migration and Hofstede's cultural dimensions, it is important to note that other variables could also be employed. For example, measures of variation in religion, language, and ethnicity composition within a country could provide additional insights. Examining the effects of these diversity variables on the Worldwide Governance Indicators using a similar methodology as employed in our study would be an interesting direction for future research. This would help to broaden our understanding of the complex relations between various forms of diversity and governance. Relations to other indicators, such as the Herfindahl--Hirschman index, which measures industry competition, could also be investigated.

In terms of methodology, it would be beneficial to employ and compare alternative models in addition to our regression model of governance. One promising approach to consider is structural equation modeling (SEM), which allows for a more comprehensive examination of the relations between variables.

\section{Conclusion}
\label{sec:con}

The traditional approach of Hofstede's cultural dimensions has been criticized for defining and measuring culture solely as ``national culture,'' failing to acknowledge the diversity within a country. To address this, we have proposed an improved method that measures the culture of immigrants in each country, thereby accounting for cultural change and current diversity. While the relation between the Hofstede's cultural dimensions and the Worldwide Governance Indicators is a common topic in the related literature, our study is unique in its use of cultural diversity indicators based on Hofstede's cultural dimensions and recent migration, which represents the primary contribution of our paper. However, as our analysis relies solely on migratory data from the past few decades, we are limited in our ability to define pre-existing minority populations. Overall, our approach provides a more comprehensive understanding of cultural diversity and its impact on governance.

We have determined the influence of culture and cultural diversity on the quality of governance of countries by admixing the data on the foreign-born populations into Hofstede's cultural dimensions as these citizens are also part of the ``national culture''. These amended cultural dimensions improved our model measuring the impact of culture on governance. We have found Power Distance, and to a lesser extent Masculinity and Uncertainty Avoidance, to have a significant negative impact on the quality of institutions. In contrast, Individualism, Long-Term Orientation, and Indulgence have a significant positive impact. 

The cultural diversity, which we defined as the standard deviation of the individual Hofstede's cultural dimensions over all citizens of the country, have varying effects on governance depending on the cultural dimension. Diversity in Uncertainty Avoidance, Long-Term Orientation, and Indulgence have positive effects on governance, albeit only for certain governance indicators. Conversely, diversity in Power Distance, Individualism, and Masculinity have negative effects on some governance indicators. Among these, the most notable correlation is the negative impact of diversity in Power Distance on Political Stability.

\section*{Funding}
\label{sec:fund}

The work of Vladimír Holý was supported by the Internal Grant Agency of the Prague University of Economics and Business under project F4/27/2020 and the personal and professional development support program of the Faculty of Informatics and Statistics, Prague University of Economics and Business.

%\bibliographystyle{myjss}
%\bibliography{library.bib,data.bib}

\begin{thebibliography}{93}
\newcommand{\enquote}[1]{``#1''}
\providecommand{\natexlab}[1]{#1}
\providecommand{\url}[1]{\texttt{#1}}
\providecommand{\urlprefix}{}
\expandafter\ifx\csname urlstyle\endcsname\relax
  \providecommand{\doi}[1]{doi:\discretionary{}{}{}#1}\else
  \providecommand{\doi}{doi:\discretionary{}{}{}\begingroup
  \urlstyle{rm}\Url}\fi
\providecommand{\eprint}[2][]{\url{#2}}

\bibitem[{Al-Marhubi(2004)}]{Al-Marhubi2004}
Al-Marhubi F (2004).
\newblock \enquote{{The Determinants of Governance: A Cross-Country Analysis}.}
\newblock \emph{Contemporary Economic Policy}, \textbf{22}(3), 394--406.
\newblock ISSN 1074-3529.
\newblock \url{https://doi.org/10.1093/cep/byh029}.

\bibitem[{Alesina \emph{et~al.}(1999)Alesina, Baqir, and
  Easterly}]{Alesina1999}
Alesina A, Baqir R, Easterly W (1999).
\newblock \enquote{{Public Goods and Ethnic Divisions}.}
\newblock \emph{Quarterly Journal of Economics}, \textbf{114}(4), 1243--1284.
\newblock ISSN 0033-5533.
\newblock \url{https://doi.org/10.1162/003355399556269}.

\bibitem[{Alesina \emph{et~al.}(2003)Alesina, Devleeschauwer, Easterly, Kurlat,
  and Wacziarg}]{Alesina2003}
Alesina A, Devleeschauwer A, Easterly W, Kurlat S, Wacziarg R (2003).
\newblock \enquote{{Fractionalization}.}
\newblock \emph{Journal of Economic Growth}, \textbf{8}(2), 155--194.
\newblock ISSN 1381-4338.
\newblock \url{https://doi.org/10.1023/a:1024471506938}.

\bibitem[{Algan and Cahuc(2010)}]{Algan2010}
Algan Y, Cahuc P (2010).
\newblock \enquote{{Inherited Trust and Growth}.}
\newblock \emph{American Economic Review}, \textbf{100}(5), 2060--2092.
\newblock ISSN 0002-8282.
\newblock \url{https://doi.org/10.1257/aer.100.5.2060}.

\bibitem[{{Awaworyi Churchill}(2017)}]{AwaworyiChurchill2017}
{Awaworyi Churchill} S (2017).
\newblock \enquote{{Fractionalization, Entrepreneurship, and the Institutional
  Environment for Entrepreneurship}.}
\newblock \emph{Small Business Economics}, \textbf{48}(3), 577--597.
\newblock ISSN 0921-898X.
\newblock \url{https://doi.org/10.1007/s11187-016-9796-8}.

\bibitem[{Baldwin and Martin(1999)}]{Baldwin1999}
Baldwin RE, Martin P (1999).
\newblock \enquote{{Two Waves of Globalisation: Superficial Similarities,
  Fundamental Differences}.}
\newblock \urlprefix\url{http://www.nber.org/papers/w6904}.

\bibitem[{Barkema and Vermeulen(1997)}]{Barkema1997}
Barkema HG, Vermeulen F (1997).
\newblock \enquote{{What Differences in the Cultural Backgrounds of Partners
  Are Detrimental for International Joint Ventures?}}
\newblock \emph{Journal of International Business Studies}, \textbf{28}(4),
  845--864.
\newblock ISSN 0047-2506.
\newblock \url{https://doi.org/10.1057/palgrave.jibs.8490122}.
\newblock
  \urlprefix\url{http://link.springer.com/10.1057/palgrave.jibs.8490122}.

\bibitem[{Beugelsdijk \emph{et~al.}(2017)Beugelsdijk, Kostova, and
  Roth}]{Beugelsdijk2017}
Beugelsdijk S, Kostova T, Roth K (2017).
\newblock \enquote{{An Overview of Hofstede-Inspired Country-Level Culture
  Research in International Business Since 2006}.}
\newblock \emph{Journal of International Business Studies}, \textbf{48}(1),
  30--47.
\newblock ISSN 0047-2506.
\newblock \url{https://doi.org/10.1057/s41267-016-0038-8}.

\bibitem[{Beugelsdijk \emph{et~al.}(2015)Beugelsdijk, Maseland, and van
  Hoorn}]{Beugelsdijk2015}
Beugelsdijk S, Maseland R, van Hoorn A (2015).
\newblock \enquote{{Are Scores on Hofstede's Dimensions of National Culture
  Stable over Time? A Cohort Analysis}.}
\newblock \emph{Global Strategy Journal}, \textbf{5}(3), 223--240.
\newblock ISSN 2042-5791.
\newblock \url{https://doi.org/10.1002/gsj.1098}.

\bibitem[{Bj{\o}rnskov(2006)}]{Bjornskov2006}
Bj{\o}rnskov C (2006).
\newblock \enquote{{The Multiple Facets of Social Capital}.}
\newblock \emph{European Journal of Political Economy}, \textbf{22}(1), 22--40.
\newblock ISSN 0176-2680.
\newblock \url{https://doi.org/10.1016/j.ejpoleco.2005.05.006}.

\bibitem[{Boeri(2010)}]{Boeri2010}
Boeri T (2010).
\newblock \enquote{{Immigration to the Land of Redistribution}.}
\newblock \emph{Economica}, \textbf{77}(308), 651--687.
\newblock ISSN 0013-0427.
\newblock \url{https://doi.org/10.1111/j.1468-0335.2010.00859.x}.

\bibitem[{Borjas(1999)}]{Borjas1999}
Borjas GJ (1999).
\newblock \enquote{{Immigration and Welfare Magnets}.}
\newblock \emph{Journal of Labor Economics}, \textbf{17}(4), 607--637.
\newblock ISSN 0734-306X.
\newblock \url{https://doi.org/10.1086/209933}.

\bibitem[{Bramson \emph{et~al.}(2017)Bramson, Grim, Singer, Berger, Sack,
  Fisher, Flocken, and Holman}]{Bramson2017}
Bramson A, Grim P, Singer DJ, Berger WJ, Sack G, Fisher S, Flocken C, Holman B
  (2017).
\newblock \enquote{{Understanding Polarization: Meanings, Measures, and Model
  Evaluation}.}
\newblock \emph{Philosophy of Science}, \textbf{84}(1), 115--159.
\newblock ISSN 0031-8248.
\newblock \url{https://doi.org/10.1086/688938}.

\bibitem[{Buller and Bell(1986)}]{Buller1986}
Buller PF, Bell CH (1986).
\newblock \enquote{{Effects of Team Building and Goal Setting on Productivity:
  A Field Experiment}.}
\newblock \emph{Academy of Management Journal}, \textbf{29}(2), 305--328.
\newblock ISSN 0001-4273.
\newblock \url{https://doi.org/10.5465/256190}.

\bibitem[{Chakraborty \emph{et~al.}(2015)Chakraborty, Thompson, and
  Yehoue}]{Chakraborty2015}
Chakraborty S, Thompson JC, Yehoue EB (2015).
\newblock \enquote{{Culture in Development}.}
\newblock \emph{The World Bank Economic Review}, \textbf{29}(Supplement 1),
  S238--S246.
\newblock ISSN 0258-6770.
\newblock \url{https://doi.org/10.1093/wber/lhv018}.

\bibitem[{Cox and Blake(1991)}]{Cox1991}
Cox TH, Blake S (1991).
\newblock \enquote{{Managing Cultural Diversity: Implications for
  Organizational Competitiveness}.}
\newblock \emph{Academy of Management Executive}, \textbf{5}(3), 45--56.
\newblock ISSN 1938-9779.
\newblock \url{https://doi.org/10.5465/ame.1991.4274465}.

\bibitem[{Dorfman and Howell(1988)}]{Dorfman1988}
Dorfman PW, Howell JP (1988).
\newblock \enquote{{Dimension of National Culture and Effective Leadership
  Patterns: Hofstede Revisited}.}
\newblock In RN~Farmer, RB~Peterson, EG~McGoun, P~Marer (Eds.), \emph{Advances
  in International Comparative Management}, Volume~3,  127--150. JAI Press,
  Greenwich.
\newblock ISBN 978-0-89232-501-6.
\newblock
  \urlprefix\url{https://books.google.com/books/about/Advances{\_}in{\_}International{\_}Comparative{\_}Ma.html?id=ELDZAAAAMAAJ}.

\bibitem[{Earley and Mosakowski(2000)}]{Earley2000}
Earley PC, Mosakowski E (2000).
\newblock \enquote{{Creating Hybrid Team Cultures: An Empirical Test of
  Transnational Team Functioning}.}
\newblock \emph{Academy of Management Journal}, \textbf{43}(1), 26--49.
\newblock ISSN 0001-4273.
\newblock \url{https://doi.org/10.2307/1556384}.

\bibitem[{Easterly(2001)}]{Easterly2001}
Easterly W (2001).
\newblock \enquote{{The Middle Class Consensus and Economic Development}.}
\newblock \emph{Journal of Economic Growth}, \textbf{6}(4), 317--335.
\newblock ISSN 1381-4338.
\newblock \url{https://doi.org/10.1023/a:1012786330095}.

\bibitem[{Easterly and Levine(1997)}]{Easterly1997}
Easterly W, Levine R (1997).
\newblock \enquote{{Africa's Growth Tragedy: Policies and Ethnic Divisions}.}
\newblock \emph{The Quarterly Journal of Economics}, \textbf{112}(4),
  1203--1250.
\newblock ISSN 0033-5533.
\newblock \url{https://doi.org/10.1162/003355300555466}.

\bibitem[{Ely and Thomas(2001)}]{Ely2001}
Ely RJ, Thomas DA (2001).
\newblock \enquote{{Cultural Diversity at Work: The Effects of Diversity
  Perspectives on Work Group Processes and Outcomes}.}
\newblock \emph{Administrative Science Quarterly}, \textbf{46}(2), 229--273.
\newblock ISSN 0001-8392.
\newblock \url{https://doi.org/10.2307/2667087}.

\bibitem[{Evan and Bolotov(2021)}]{Evan2021}
Evan T, Bolotov I (2021).
\newblock \enquote{{Measuring Mancur Olson: What is the Influence of Culture,
  Institutions and Policies on Economic Development?}}
\newblock \emph{Prague Economic Papers}, \textbf{30}(3), 290--315.
\newblock ISSN 1210-0455.
\newblock \url{https://doi.org/10.18267/j.pep.770}.

\bibitem[{Fearon and Laitin(1996)}]{Fearon1996}
Fearon JD, Laitin DD (1996).
\newblock \enquote{{Explaining Interethnic Cooperation}.}
\newblock \emph{American Political Science Review}, \textbf{90}(4), 715--735.
\newblock ISSN 0003-0554.
\newblock \url{https://doi.org/10.2307/2945838}.

\bibitem[{Fern{\'{a}}ndez(2010)}]{Fernandez2010}
Fern{\'{a}}ndez R (2010).
\newblock \enquote{{Does Culture Matter?}}
\newblock \urlprefix\url{http://www.nber.org/papers/w16277}.

\bibitem[{Gorodnichenko and Roland(2011)}]{Gorodnichenko2011}
Gorodnichenko Y, Roland G (2011).
\newblock \enquote{{Individualism, Innovation, and Long-Run Growth}.}
\newblock \emph{Proceedings of the National Academy of Sciences of the United
  States of America}, \textbf{108}(Supplement 4), 21316--21319.
\newblock ISSN 0027-8424.
\newblock \url{https://doi.org/10.1073/pnas.1101933108}.

\bibitem[{Grafton \emph{et~al.}(2002)Grafton, Knowles, and Owen}]{Grafton2002}
Grafton RQ, Knowles S, Owen PD (2002).
\newblock \enquote{{Social Divergence and Productivity: Making a Connection}.}
\newblock In K~Banting, A~Sharpe, F~St-Hilaire (Eds.), \emph{The Review of
  Economic Performance and Social Progress: Towards a Social Understanding of
  Productivity},  203--224. Institute for Research on Public Policy.
\newblock ISBN 978-0-88645-198-1.
\newblock \urlprefix\url{https://books.google.com/books?id=9GyuAAAAIAAJ}.

\bibitem[{Green and Urquhart(1976)}]{Green1976}
Green A, Urquhart MC (1976).
\newblock \enquote{{Factor and Commodity Flows in the International Economy of
  1870-1914: A Multi-Country View}.}
\newblock \emph{The Journal of Economic History}, \textbf{36}(1), 217--252.
\newblock ISSN 0022-0507.
\newblock \url{https://doi.org/10.1017/S0022050700094651}.

\bibitem[{Greene(2018)}]{Greene2018}
Greene WH (2018).
\newblock \emph{{Econometric Analysis}}.
\newblock Eight Edition. Pearson, New York.
\newblock ISBN 978-0-13-446136-6.
\newblock
  \urlprefix\url{https://www.pearson.com/us/higher-education/program/Greene-Econometric-Analysis-8th-Edition/PGM334862.html}.

\bibitem[{Greve and Salaff(2003)}]{Greve2003}
Greve A, Salaff JW (2003).
\newblock \enquote{{Social Networks and Entrepreneurship}.}
\newblock \emph{Entrepreneurship: Theory and Practice}, \textbf{28}(1), 1--22.
\newblock ISSN 1042-2587.
\newblock \url{https://doi.org/10.1111/1540-8520.00029}.

\bibitem[{Gr{\"{o}}zinger \emph{et~al.}(2017)Gr{\"{o}}zinger, Langholz-Kaiser,
  and Richter}]{Grozinger2017}
Gr{\"{o}}zinger G, Langholz-Kaiser M, Richter D (2017).
\newblock \enquote{{Regional Innovation and Diversity: Effects of Cultural
  Diversity, Milieu Affiliation and Qualification Levels on Regional Patent
  Outputs}.}
\newblock \emph{Management Revue}, \textbf{28}(2), 149--174.
\newblock ISSN 0935-9915.
\newblock \url{https://doi.org/10.5771/0935-9915-2017-2-149}.

\bibitem[{Harvey and Kou(2013)}]{Harvey2013a}
Harvey S, Kou CY (2013).
\newblock \enquote{{Collective Engagement in Creative Tasks: The Role of
  Evaluation in the Creative Process in Groups}.}
\newblock \emph{Administrative Science Quarterly}, \textbf{58}(3), 346--386.
\newblock ISSN 0001-8392.
\newblock \url{https://doi.org/10.1177/0001839213498591}.

\bibitem[{Hofstede \emph{et~al.}(2010)Hofstede, Hofstede, and
  Minkov}]{Hofstede2010}
Hofstede G, Hofstede GJ, Minkov M (2010).
\newblock \emph{{Cultures and Organizations: Software of the Mind}}, Volume~10.
\newblock Third Edition. McGraw-Hill.
\newblock ISBN 978-0-07-177015-6.
\newblock
  \urlprefix\url{https://www.mhprofessional.com/9780071664189-usa-cultures-and-organizations-software-of-the-mind-third-edition-group}.

\bibitem[{{Hofstede Insights}(2021)}]{Hofstede2021}
{Hofstede Insights} (2021).
\newblock \enquote{{National Culture}.}
\newblock
  \urlprefix\url{https://www.hofstede-insights.com/models/national-culture/}.

\bibitem[{Hol{\'{y}} and Evan(2022)}]{Holy2022d}
Hol{\'{y}} V, Evan T (2022).
\newblock \enquote{{The Role of a Nation's Culture in the Country's Governance:
  Stochastic Frontier Analysis}.}
\newblock \emph{Central European Journal of Operations Research},
  \textbf{30}(2), 507--520.
\newblock ISSN 1435-246X.
\newblock \url{https://doi.org/10.1007/s10100-021-00754-5}.

\bibitem[{House \emph{et~al.}(2004)House, Hanges, Javidan, Dorfman, and
  Gupta}]{House2004}
House RJ, Hanges PJ, Javidan M, Dorfman PW, Gupta V (2004).
\newblock \emph{{Culture, Leadership, and Organizations: The GLOBE Study of 62
  Societies}}.
\newblock Sage Publications, Thousand Oaks.
\newblock ISBN 978-0-7619-2401-2.
\newblock
  \urlprefix\url{https://us.sagepub.com/en-us/nam/culture-leadership-and-organizations/book226013}.

\bibitem[{Inglehart \emph{et~al.}(2004)Inglehart, Basanez, Dies-Medriano,
  Halman, and Luijkx}]{Inglehart2004}
Inglehart R, Basanez M, Dies-Medriano J, Halman L, Luijkx R (2004).
\newblock \emph{{Human Beliefs and Values: A Cross-cultural Sourcebook Based on
  the 1999-2002 Values Surveys}}.
\newblock Siglo XXI, Mexico City.
\newblock ISBN 978-968-23-2502-1.

\bibitem[{Jones(2007)}]{Jones2007}
Jones W (2007).
\newblock \enquote{{Personal Information Management}.}
\newblock \emph{Annual Review of Information Science and Technology},
  \textbf{41}(1), 453--504.
\newblock ISSN 0066-4200.
\newblock \url{https://doi.org/10.1002/aris.2007.1440410117}.

\bibitem[{Kaufmann \emph{et~al.}(2011)Kaufmann, Kraay, and
  Mastruzzi}]{Kaufmann2011}
Kaufmann D, Kraay A, Mastruzzi M (2011).
\newblock \enquote{{The Worldwide Governance Indicators: Methodology and
  Analytical Issues}.}
\newblock \emph{Hague Journal on the Rule of Law}, \textbf{3}(2), 220--246.
\newblock ISSN 1876-4045.
\newblock \url{https://doi.org/10.1017/s1876404511200046}.

\bibitem[{Khayesi \emph{et~al.}(2014)Khayesi, George, and
  Antonakis}]{Khayesi2014}
Khayesi JNO, George G, Antonakis J (2014).
\newblock \enquote{{Kinship in Entrepreneur Networks: Performance Effects of
  Resource Assembly in Africa}.}
\newblock \emph{Entrepreneurship: Theory and Practice}, \textbf{38}(6),
  1323--1342.
\newblock ISSN 1042-2587.
\newblock \url{https://doi.org/10.1111/etap.12127}.

\bibitem[{Kirkman \emph{et~al.}(2006)Kirkman, Lowe, and Gibson}]{Kirkman2006}
Kirkman BL, Lowe KB, Gibson CB (2006).
\newblock \enquote{{A Quarter Century of "Culture's Consequences": A Review of
  Empirical Research Incorporating Hofstede's Cultural Values Framework}.}
\newblock \emph{Journal of International Business Studies}, \textbf{37}(3),
  285--320.
\newblock ISSN 0047-2506.
\newblock \url{https://doi.org/10.1057/palgrave.jibs.8400202}.

\bibitem[{Koenig \emph{et~al.}(2012)Koenig, Wagener, Ifo, Aper, and
  Inance}]{Koenig2012}
Koenig T, Wagener A, Ifo CES, Aper WOP, Inance CAPUF (2012).
\newblock \enquote{{Culture and Tax Structures}.}
\newblock \urlprefix\url{https://ssrn.com/abstract=2016072}.

\bibitem[{Krain(1998)}]{Krain1998}
Krain M (1998).
\newblock \enquote{{Contemporary Democracies Revisited: Democracy, Political
  Violence, and Event Count Models}.}
\newblock \emph{Comparative Political Studies}, \textbf{31}(2), 139--164.
\newblock ISSN 0010-4140.
\newblock \url{https://doi.org/10.1177/0010414098031002001}.

\bibitem[{Laitin and Posner(2001)}]{Laitin2001}
Laitin D, Posner D (2001).
\newblock \enquote{{The Implications of Constructivism for Constructing Ethnic
  Fractionalization Indices}.}
\newblock \emph{Comparative Politics Newsletter}, \textbf{12}(1), 13--17.
\newblock
  \urlprefix\url{https://www.comparativepoliticsnewsletter.org/wp-content/uploads/2021/04/2001{\_}winter.pdf}.

\bibitem[{Laitin \emph{et~al.}(2012)Laitin, Moortgat, and
  Robinson}]{Laitin2012}
Laitin DD, Moortgat J, Robinson AL (2012).
\newblock \enquote{{Geographic Axes and the Persistence of Cultural
  Diversity}.}
\newblock \emph{Proceedings of the National Academy of Sciences of the United
  States of America}, \textbf{109}(26), 10263--10268.
\newblock ISSN 0027-8424.
\newblock \url{https://doi.org/10.1073/pnas.1205338109}.

\bibitem[{Landes(1998)}]{Landes1998}
Landes DS (1998).
\newblock \emph{{The Wealth and Poverty of Nations: Why Some Are So Rich and
  Some So Poor}}.
\newblock First Edition. W. W. Norton {\&} Company, New York.
\newblock ISBN 978-0-393-04017-3.
\newblock
  \urlprefix\url{https://wwnorton.com/books/The-Wealth-and-Poverty-of-Nations/}.

\bibitem[{Langbein and Knack(2010)}]{Langbein2010}
Langbein L, Knack S (2010).
\newblock \enquote{{The Worldwide Governance Indicators: Six, One, or None?}}
\newblock \emph{Journal of Development Studies}, \textbf{46}(2), 350--370.
\newblock ISSN 0022-0388.
\newblock \url{https://doi.org/10.1080/00220380902952399}.

\bibitem[{Lappi-Sepp{\"{a}}l{\"{a}} and Lehti(2014)}]{LappiSeppala2015}
Lappi-Sepp{\"{a}}l{\"{a}} T, Lehti M (2014).
\newblock \enquote{{Cross-Comparative Perspectives on Global Homicide Trends}.}
\newblock \emph{Crime and Justice}, \textbf{43}(1), 135--230.
\newblock ISSN 0192-3234.
\newblock \url{https://doi.org/10.1086/677979}.

\bibitem[{Lian and Oneal(1997)}]{Lian1997}
Lian B, Oneal JR (1997).
\newblock \enquote{{Cultural Diversity and Economic Development: A
  Cross-National Study of 98 Countries, 1960-1985}.}
\newblock \emph{Economic Development and Cultural Change}, \textbf{46}(1),
  61--77.
\newblock ISSN 0013-0079v.
\newblock \url{https://doi.org/10.1086/452321}.

\bibitem[{Lloyd(1980)}]{Lloyd1980}
Lloyd WF (1980).
\newblock \enquote{{W. F. Lloyd on the Checks to Population}.}
\newblock \emph{Population and Development Review}, \textbf{6}(3), 473--496.
\newblock ISSN 0098-7921.
\newblock \url{https://doi.org/10.2307/1972412}.

\bibitem[{Luttmer and Singhal(2011)}]{Luttmer2011}
Luttmer EFP, Singhal M (2011).
\newblock \enquote{{Culture, Context, and the Taste for Redistribution}.}
\newblock \emph{American Economic Journal: Economic Policy}, \textbf{3}(1),
  157--179.
\newblock ISSN 1945-774X.
\newblock \url{https://doi.org/10.1257/pol.3.1.157}.

\bibitem[{Maseland(2013)}]{Maseland2013}
Maseland R (2013).
\newblock \enquote{{Parasitical Cultures? The Cultural Origins of Institutions
  and Development}.}
\newblock \emph{Journal of Economic Growth}, \textbf{18}(2), 109--136.
\newblock ISSN 1381-4338.
\newblock \url{https://doi.org/10.1007/s10887-013-9089-x}.

\bibitem[{Matei and Abrudan(2018)}]{Matei2018}
Matei MC, Abrudan MM (2018).
\newblock \enquote{{Are National Cultures Changing? Evidence from the World
  Values Survey}.}
\newblock \emph{Procedia - Social and Behavioral Sciences}, \textbf{238},
  657--664.
\newblock ISSN 1877-0428.
\newblock \url{https://doi.org/10.1016/j.sbspro.2018.04.047}.

\bibitem[{McSweeney(2002)}]{McSweeney2002}
McSweeney B (2002).
\newblock \enquote{{Hofstede's Model of National Cultural Differences and Their
  Consequences: A Triumph of Faith - A Failure of Analysis}.}
\newblock \emph{Human Relations}, \textbf{55}(1), 89--118.
\newblock ISSN 0018-7267.
\newblock \url{https://doi.org/10.1177/0018726702551004}.

\bibitem[{Meierrieks and Gries(2013)}]{Meierrieks2013}
Meierrieks D, Gries T (2013).
\newblock \enquote{{Causality Between Terrorism and Economic Growth}.}
\newblock \emph{Journal of Peace Research}, \textbf{50}(1), 91--104.
\newblock ISSN 0022-3433.
\newblock \url{https://doi.org/10.1177/0022343312445650}.

\bibitem[{{Migration Data Portal}(2021)}]{InternationalMigrantStock2021}
{Migration Data Portal} (2021).
\newblock \enquote{{International Migrant Stock}.}
\newblock
  \urlprefix\url{https://www.migrationdataportal.org/themes/international-migrant-stocks}.

\bibitem[{Miguel \emph{et~al.}(2004)Miguel, Satyanath, and
  Sergenti}]{Miguel2004}
Miguel E, Satyanath S, Sergenti E (2004).
\newblock \enquote{{Economic Shocks and Civil Conflict: An Instrumental
  Variables Approach}.}
\newblock \emph{Journal of Political Economy}, \textbf{112}(4), 725--753.
\newblock ISSN 0022-3808.
\newblock \url{https://doi.org/10.1086/421174}.

\bibitem[{Mill(1862)}]{Mill1862}
Mill JS (1862).
\newblock \emph{{Principles of Political Economy}}.
\newblock Fifth Edition. Parker, Son, and Bourn, West Strand, London.

\bibitem[{Montgomery(1991)}]{Montgomery1991}
Montgomery JD (1991).
\newblock \enquote{{Social Networks and Labor-Market Outcomes: Toward an
  Economic Analysis}.}
\newblock \emph{American Economic Review}, \textbf{81}(5), 1408--1418.
\newblock ISSN 0002-8282.
\newblock \url{https://doi.org/10.2307/2006929}.

\bibitem[{Nettle(2000)}]{Nettle2000}
Nettle D (2000).
\newblock \enquote{{Linguistic Fragmentation and the Wealth of Nations: The
  Fishman-Pool Hypothesis Reexamined}.}
\newblock \emph{Economic Development and Cultural Change}, \textbf{48}(2),
  335--348.
\newblock ISSN 0013-0079.
\newblock \url{https://doi.org/10.1086/452461}.

\bibitem[{Nettle \emph{et~al.}(2007)Nettle, Grace, Choisy, Cornell,
  Gu{\'{e}}gan, and Hochberg}]{Nettle2007}
Nettle D, Grace JB, Choisy M, Cornell HV, Gu{\'{e}}gan JF, Hochberg ME (2007).
\newblock \enquote{{Cultural Diversity, Economic Development and Societal
  Instability}.}
\newblock \emph{PLoS ONE}, \textbf{2}(9), e929/1--e929/5.
\newblock ISSN 1932-6203.
\newblock \url{https://doi.org/10.1371/journal.pone.0000929}.

\bibitem[{Niebuhr(2010)}]{Niebuhr2010}
Niebuhr A (2010).
\newblock \enquote{{Migration and Innovation: Does Cultural Diversity Matter
  for Regional R{\&}D Activity?}}
\newblock \emph{Papers in Regional Science}, \textbf{89}(3), 563--585.
\newblock ISSN 1056-8190.
\newblock \url{https://doi.org/10.1111/j.1435-5957.2009.00271.x}.

\bibitem[{North(1990)}]{North1990}
North DC (1990).
\newblock \emph{{Institutions, Institutional Change and Economic Performance}}.
\newblock Cambridge University Press, Cambridge.
\newblock ISBN 978-0-511-80867-8.
\newblock \url{https://doi.org/10.1017/cbo9780511808678}.

\bibitem[{North(1992)}]{North1992a}
North DC (1992).
\newblock \enquote{{Institutions and Economic Theory}.}
\newblock \emph{The American Economist}, \textbf{36}(1), 3--6.
\newblock ISSN 0569-4345.
\newblock \url{https://doi.org/10.1177/056943459203600101}.

\bibitem[{North(1994)}]{North1994}
North DC (1994).
\newblock \enquote{{Economic Performance Through Time}.}
\newblock \emph{American Economic Review}, \textbf{84}(3), 359--368.
\newblock ISSN 0002-8282.
\newblock \url{https://doi.org/10.2307/2118057}.

\bibitem[{North and Thomas(1977)}]{North1977}
North DC, Thomas RP (1977).
\newblock \enquote{{The First Economic Revolution}.}
\newblock \emph{The Economic History Review}, \textbf{30}(2), 229--241.
\newblock ISSN 0013-0117.
\newblock \url{https://doi.org/10.2307/2595144}.

\bibitem[{Olson(1996)}]{Olson1996}
Olson M (1996).
\newblock \enquote{{Distinguished Lecture on Economics in Government: Big Bills
  Left on the Sidewalk: Why Some Nations are Rich, and Others Poor}.}
\newblock \emph{Journal of Economic Perspectives}, \textbf{10}(2), 3--24.
\newblock ISSN 0895-3309.
\newblock \url{https://doi.org/10.1257/jep.10.2.3}.

\bibitem[{{\O}stergaard \emph{et~al.}(2011){\O}stergaard, Timmermans, and
  Kristinsson}]{Ostergaard2011}
{\O}stergaard CR, Timmermans B, Kristinsson K (2011).
\newblock \enquote{{Does a Different View Create Something New? The Effect of
  Employee Diversity on Innovation}.}
\newblock \emph{Research Policy}, \textbf{40}(3), 500--509.
\newblock ISSN 0048-7333.
\newblock \url{https://doi.org/10.1016/j.respol.2010.11.004}.

\bibitem[{Ottaviano and Peri(2006)}]{Ottaviano2006}
Ottaviano GIP, Peri G (2006).
\newblock \enquote{{The Economic Value of Cultural Diversity: Evidence from US
  Cities}.}
\newblock \emph{Journal of Economic Geography}, \textbf{6}(1), 9--44.
\newblock ISSN 1468-2702.
\newblock \url{https://doi.org/10.1093/jeg/lbi002}.

\bibitem[{Outreville(2018)}]{Outreville2018}
Outreville JF (2018).
\newblock \enquote{{Culture and Life Insurance Ownership: Is It an Issue?}}
\newblock \emph{Journal of Insurance Issues}, \textbf{41}(2), 168--192.
\newblock ISSN 1531-6076.
\newblock \urlprefix\url{https://www.jstor.org/stable/26505753}.

\bibitem[{Pelled \emph{et~al.}(1999)Pelled, Ledford, and Mohrman}]{Pelled1999}
Pelled LH, Ledford GE, Mohrman SA (1999).
\newblock \enquote{{Demographic Dissimilarity and Workplace Inclusion}.}
\newblock \emph{Journal of Management Studies}, \textbf{36}(7), 1013--1031.
\newblock ISSN 0022-2380.
\newblock \url{https://doi.org/10.1111/1467-6486.00168}.

\bibitem[{Pool(1972)}]{Pool1972}
Pool J (1972).
\newblock \enquote{{National Development and Language Diversity}.}
\newblock In JA~Fishman (Ed.), \emph{Advances in the Sociology of Language},
  Volume~2, Chapter~11,  213--230. De Gruyter, The Hague.
\newblock ISBN 978-90-279-2302-8.
\newblock \url{https://doi.org/10.1515/9783110880434-011}.

\bibitem[{Pryor(2007)}]{Pryor2007}
Pryor FL (2007).
\newblock \enquote{{Culture and Economic Systems}.}
\newblock \emph{American Journal of Economics and Sociology}, \textbf{66}(4),
  817--855.
\newblock ISSN 0002-9246.
\newblock \url{https://doi.org/10.1111/j.1536-7150.2007.00541.x}.

\bibitem[{Putnam(2007)}]{Putnam2007}
Putnam RD (2007).
\newblock \enquote{{E Pluribus Unum: Diversity and Community in the
  Twenty-First Century: The 2006 Johan Skytte Prize Lecture}.}
\newblock \emph{Scandinavian Political Studies}, \textbf{30}(2), 137--174.
\newblock ISSN 0080-6757.
\newblock \url{https://doi.org/10.1111/j.1467-9477.2007.00176.x}.

\bibitem[{Rejchrt and Higgs(2015)}]{Rejchrt2015}
Rejchrt P, Higgs M (2015).
\newblock \enquote{{When in Rome: How Non-Domestic Companies Listed in the UK
  May Not Comply with Accepted Norms and Principles of Good Corporate
  Governance. Does Home Market Culture Explain These Corporate Behaviours and
  Attitudes to Compliance?}}
\newblock \emph{Journal of Business Ethics}, \textbf{129}(1), 131--159.
\newblock ISSN 0167-4544.
\newblock \url{https://doi.org/10.1007/s10551-014-2151-6}.

\bibitem[{Richard \emph{et~al.}(2004)Richard, Barnett, Dwyer, and
  Chadwick}]{Richard2004}
Richard OC, Barnett T, Dwyer S, Chadwick K (2004).
\newblock \enquote{{Cultural Diversity in Management, Firm Performance, and the
  Moderating Role of Entrepreneurial Orientation Dimensions}.}
\newblock \emph{Academy of Management Journal}, \textbf{47}(2), 255--266.
\newblock ISSN 0001-4273.
\newblock \url{https://doi.org/10.5465/20159576}.

\bibitem[{Schwartz(1999)}]{Schwartz1999}
Schwartz SH (1999).
\newblock \enquote{{A Theory of Cultural Values and Some Implications for
  Work}.}
\newblock \emph{Applied Psychology}, \textbf{48}(1), 23--47.
\newblock ISSN 0269-994X.
\newblock \url{https://doi.org/10.1080/026999499377655}.

\bibitem[{Schwartz \emph{et~al.}(2012)Schwartz, Cieciuch, Vecchione, Davidov,
  Fischer, Beierlein, Ramos, Verkasalo, L{\"{o}}nnqvist, Demirutku,
  Dirilen-Gumus, and Konty}]{Schwartz2012}
Schwartz SH, Cieciuch J, Vecchione M, Davidov E, Fischer R, Beierlein C, Ramos
  A, Verkasalo M, L{\"{o}}nnqvist JE, Demirutku K, Dirilen-Gumus O, Konty M
  (2012).
\newblock \enquote{{Refining the Theory of Basic Individual Values}.}
\newblock \emph{Journal of Personality and Social Psychology}, \textbf{103}(4),
  663--688.
\newblock \url{https://doi.org/10.1037/a0029393}.

\bibitem[{Stahl \emph{et~al.}(2010)Stahl, Maznevski, Voigt, and
  Jonsen}]{Stahl2010}
Stahl GK, Maznevski ML, Voigt A, Jonsen K (2010).
\newblock \enquote{{Unraveling the Effects of Cultural Diversity in Teams: A
  Meta-Analysis of Research on Multicultural Work Groups}.}
\newblock \emph{Journal of International Business Studies}, \textbf{41}(4),
  690--709.
\newblock ISSN 0047-2506.
\newblock \url{https://doi.org/10.1057/jibs.2009.85}.

\bibitem[{Tang and Koveos(2008)}]{Tang2008}
Tang L, Koveos PE (2008).
\newblock \enquote{{A Framework to Update Hofstede's Cultural Value Indices:
  Economic Dynamics and Institutional Stability}.}
\newblock \emph{Journal of International Business Studies}, \textbf{39}(6),
  1045--1063.
\newblock ISSN 0047-2506.
\newblock \url{https://doi.org/10.1057/palgrave.jibs.8400399}.

\bibitem[{Taras \emph{et~al.}(2016)Taras, Steel, and Kirkman}]{Taras2016}
Taras V, Steel P, Kirkman BL (2016).
\newblock \enquote{{Does Country Equate with Culture? Beyond Geography in the
  Search for Cultural Boundaries}.}
\newblock \emph{Management International Review}, \textbf{56}(4), 455--487.
\newblock ISSN 0938-8249.
\newblock \url{https://doi.org/10.1007/s11575-016-0283-x}.

\bibitem[{Taylor(2000)}]{Taylor2000}
Taylor JB (2000).
\newblock \enquote{{Low Inflation, Pass-Through, and the Pricing Power of
  Firms}.}
\newblock \emph{European Economic Review}, \textbf{44}(7), 1389--1408.
\newblock ISSN 0014-2921.
\newblock \url{https://doi.org/10.1016/s0014-2921(00)00037-4}.

\bibitem[{Taylor and Wilson(2012)}]{Taylor2012}
Taylor MZ, Wilson S (2012).
\newblock \enquote{{Does Culture Still Matter?: The Effects of Individualism on
  National Innovation Rates}.}
\newblock \emph{Journal of Business Venturing}, \textbf{27}(2), 234--247.
\newblock ISSN 0883-9026.
\newblock \url{https://doi.org/10.1016/j.jbusvent.2010.10.001}.

\bibitem[{{United Nations Development
  Programme}(2021)}]{HumanDevelopmentIndex2021}
{United Nations Development Programme} (2021).
\newblock \enquote{{Human Development Index (HDI)}.}
\newblock
  \urlprefix\url{https://hdr.undp.org/data-center/human-development-index}.

\bibitem[{Vance and Paik(2014)}]{Vance2015}
Vance CM, Paik Y (2014).
\newblock \emph{{Managing a Global Workforce: Challenges and Opportunities in
  International Human Resource Management}}.
\newblock Third Edition. Routledge, New York.
\newblock ISBN 978-1-315-71964-1.
\newblock \url{https://doi.org/10.4324/9781315719641}.

\bibitem[{Walker and Poe(2002)}]{Walker2002}
Walker S, Poe SC (2002).
\newblock \enquote{{Does Cultural Diversity Affect Countries' Respect for Human
  Rights?}}
\newblock \emph{Human Rights Quarterly}, \textbf{24}(1), 237--263.
\newblock ISSN 0275-0392.
\newblock \url{https://doi.org/10.1353/hrq.2002.0016}.

\bibitem[{Watson \emph{et~al.}(1993)Watson, Kumar, and Michaelsen}]{Watson1993}
Watson WE, Kumar K, Michaelsen LK (1993).
\newblock \enquote{{Cultural Diversity's Impact on Interaction Process and
  Performance: Comparing Homogeneous and Diverse Task Groups}.}
\newblock \emph{Academy of Management Journal}, \textbf{36}(3), 590--602.
\newblock ISSN 0001-4273.
\newblock \url{https://doi.org/10.5465/256593}.

\bibitem[{Williams and {O'Reilly III}(1998)}]{Williams1998}
Williams KY, {O'Reilly III} Ca (1998).
\newblock \enquote{{Demography and Diversity in Organizations: A Review of 40
  Years of Research}.}
\newblock \emph{Research in Organizational Behavior}, \textbf{20}, 77--140.
\newblock ISSN 0191-3085.
\newblock \urlprefix\url{https://www.researchgate.net/publication/234022034}.

\bibitem[{Williams and McGuire(2010)}]{Williams2010}
Williams LK, McGuire SJ (2010).
\newblock \enquote{{Economic Creativity and Innovation Implementation: The
  Entrepreneurial Drivers of Growth? Evidence from 63 Countries}.}
\newblock \emph{Small Business Economics}, \textbf{34}(4), 391--412.
\newblock ISSN 0921-898X.
\newblock \url{https://doi.org/10.1007/s11187-008-9145-7}.

\bibitem[{{World Bank}(2021)}]{WGI2021}
{World Bank} (2021).
\newblock \enquote{{Worldwide Governance Indicators}.}
\newblock \urlprefix\url{https://info.worldbank.org/governance/wgi/}.

\bibitem[{Wu \emph{et~al.}(2012)Wu, Li, and Selover}]{Wu2012}
Wu J, Li S, Selover DD (2012).
\newblock \enquote{{Foreign Direct Investment vs. Foreign Portfolio Investment:
  The Effect of the Governance Environment}.}
\newblock \emph{Management International Review}, \textbf{52}(5), 643--670.
\newblock ISSN 0938-8249.
\newblock \url{https://doi.org/10.1007/s11575-011-0121-0}.

\bibitem[{Zak and Knack(2001)}]{Zak2001}
Zak PJ, Knack S (2001).
\newblock \enquote{{Trust and Growth}.}
\newblock \emph{The Economic Journal}, \textbf{111}(470), 295--321.
\newblock ISSN 0013-0133.
\newblock \url{https://doi.org/10.1111/1468-0297.00609}.

\bibitem[{Zhan \emph{et~al.}(2015)Zhan, Bendapudi, and Hong}]{Zhan2015}
Zhan S, Bendapudi N, Hong YY (2015).
\newblock \enquote{{Re-Examining Diversity as a Double-Edged Sword for
  Innovation Process}.}
\newblock \emph{Journal of Organizational Behavior}, \textbf{36}(7),
  1026--1049.
\newblock ISSN 0894-3796.
\newblock \url{https://doi.org/10.1002/job.2027}.

\bibitem[{Zhao \emph{et~al.}(2016)Zhao, Kwon, and Yang}]{Zhao2016}
Zhao HY, Kwon JW, Yang OS (2016).
\newblock \enquote{{Updating Hofstede's Cultural Model and Tracking Changes in
  Cultural Indices}.}
\newblock \emph{Journal of International Trade {\&} Commerce}, \textbf{12}(5),
  85--106.
\newblock ISSN 1738-8112.
\newblock \url{https://doi.org/10.16980/jitc.12.5.201610.85}.

\end{thebibliography}

\appendix

\section{Descriptive Statistics}
\label{app:country}
\captionsetup{width=15cm}
\begin{landscape}
\footnotesize
\begin{longtable}{l *{18}{r} }
\caption{Average values of Worldwide Governance Indicators, Hofstede’s cultural levels, and Hofstede’s cultural diversities over 2000, 2005, 2010, 2015, and 2020.}
\label{tab:country} \\
\toprule
& \multicolumn{6}{c}{Worldwide Governance Indicators} & \multicolumn{6}{c}{Hofstede’s Cultural Levels} & \multicolumn{6}{c}{Hofstede’s Cultural Diversities} \\
\cmidrule(l{3pt}r{3pt}){2-7} \cmidrule(l{3pt}r{3pt}){8-13} \cmidrule(l{3pt}r{3pt}){14-19}
Country & VA & PV & GE & RQ & RL & CC & PDI & IDV & MAS & UAI & LTO & IVR & PDI & IDV & MAS & UAI & LTO & IVR \\
\midrule
\endfirsthead
\toprule
& \multicolumn{6}{c}{Worldwide Governance Indicators} & \multicolumn{6}{c}{Hofstede’s Cultural Levels} & \multicolumn{6}{c}{Hofstede’s Cultural Diversities} \\
\cmidrule(l{3pt}r{3pt}){2-7} \cmidrule(l{3pt}r{3pt}){8-13} \cmidrule(l{3pt}r{3pt}){14-19}
Country & VA & PV & GE & RQ & RL & CC & PDI & IDV & MAS & UAI & LTO & IVR & PDI & IDV & MAS & UAI & LTO & IVR \\
\midrule
\endhead
\bottomrule
\endfoot
Albania & 0.02 & -0.16 & -0.36 & 0.01 & -0.57 & -0.64 & 0.89 & 0.20 & 0.80 & 0.70 & 0.61 & 0.16 & 0.05 & 0.04 & 0.03 & 0.04 & 0.02 & 0.05 \\ 
  Algeria & -0.96 & -1.11 & -0.59 & -0.94 & -0.88 & -0.65 & 0.80 & 0.35 & 0.35 & 0.70 & 0.26 & 0.32 & 0.01 & 0.01 & 0.01 & 0.01 & 0.01 & 0.01 \\ 
  Angola & -1.16 & -0.83 & -1.18 & -1.18 & -1.28 & -1.29 & 0.83 & 0.18 & 0.20 & 0.60 & 0.15 & 0.82 & 0.02 & 0.02 & 0.03 & 0.02 & 0.03 & 0.04 \\ 
  Argentina & 0.41 & 0.00 & -0.11 & -0.51 & -0.51 & -0.32 & 0.50 & 0.45 & 0.55 & 0.86 & 0.21 & 0.61 & 0.04 & 0.06 & 0.04 & 0.02 & 0.04 & 0.04 \\ 
  Armenia & -0.46 & -0.30 & -0.22 & 0.15 & -0.37 & -0.53 & 0.84 & 0.22 & 0.50 & 0.87 & 0.60 & 0.25 & 0.03 & 0.03 & 0.02 & 0.04 & 0.05 & 0.03 \\ 
  Australia & 1.41 & 0.97 & 1.70 & 1.71 & 1.74 & 1.89 & 0.42 & 0.80 & 0.59 & 0.50 & 0.28 & 0.65 & 0.14 & 0.21 & 0.07 & 0.11 & 0.16 & 0.14 \\ 
  Austria & 1.38 & 1.02 & 1.71 & 1.48 & 1.83 & 1.67 & 0.20 & 0.53 & 0.75 & 0.71 & 0.59 & 0.59 & 0.22 & 0.09 & 0.11 & 0.07 & 0.07 & 0.11 \\ 
  Azerbaijan & -1.29 & -0.73 & -0.59 & -0.49 & -0.84 & -1.11 & 0.85 & 0.22 & 0.50 & 0.88 & 0.61 & 0.22 & 0.02 & 0.02 & 0.01 & 0.01 & 0.02 & 0.01 \\ 
  Bangladesh & -0.47 & -1.23 & -0.75 & -0.91 & -0.80 & -1.07 & 0.80 & 0.20 & 0.55 & 0.60 & 0.47 & 0.20 & 0.02 & 0.02 & 0.01 & 0.02 & 0.02 & 0.02 \\ 
  Belarus & -1.51 & -0.04 & -0.80 & -1.19 & -1.05 & -0.47 & 0.94 & 0.26 & 0.22 & 0.94 & 0.81 & 0.15 & 0.04 & 0.04 & 0.05 & 0.02 & 0.02 & 0.02 \\ 
  Belgium & 1.36 & 0.78 & 1.56 & 1.28 & 1.38 & 1.47 & 0.64 & 0.71 & 0.53 & 0.90 & 0.76 & 0.54 & 0.06 & 0.10 & 0.06 & 0.10 & 0.14 & 0.08 \\ 
  Bhutan & -0.50 & 0.90 & 0.50 & -0.56 & 0.34 & 1.08 & 0.93 & 0.52 & 0.33 & 0.29 & 0.55 & 0.30 & 0.05 & 0.02 & 0.06 & 0.04 & 0.02 & 0.02 \\ 
  Bolivia & -0.02 & -0.47 & -0.55 & -0.67 & -0.89 & -0.62 & 0.78 & 0.10 & 0.42 & 0.87 & 0.25 & 0.46 & 0.02 & 0.04 & 0.02 & 0.02 & 0.02 & 0.02 \\ 
  Bosnia and Herzegovina & -0.08 & -0.53 & -0.77 & -0.32 & -0.41 & -0.43 & 0.90 & 0.22 & 0.48 & 0.87 & 0.70 & 0.44 & 0.02 & 0.02 & 0.02 & 0.01 & 0.02 & 0.02 \\ 
  Brazil & 0.41 & -0.16 & -0.13 & 0.04 & -0.18 & -0.16 & 0.69 & 0.38 & 0.49 & 0.76 & 0.44 & 0.59 & 0.01 & 0.01 & 0.01 & 0.01 & 0.02 & 0.01 \\ 
  Bulgaria & 0.46 & 0.28 & 0.05 & 0.51 & -0.10 & -0.17 & 0.70 & 0.30 & 0.40 & 0.85 & 0.69 & 0.16 & 0.02 & 0.02 & 0.02 & 0.02 & 0.02 & 0.03 \\ 
  Burkina Faso & -0.24 & -0.46 & -0.60 & -0.29 & -0.46 & -0.19 & 0.67 & 0.15 & 0.48 & 0.53 & 0.26 & 0.18 & 0.03 & 0.01 & 0.02 & 0.02 & 0.02 & 0.03 \\ 
  Canada & 1.47 & 1.06 & 1.80 & 1.62 & 1.73 & 1.91 & 0.43 & 0.72 & 0.52 & 0.49 & 0.38 & 0.62 & 0.13 & 0.17 & 0.07 & 0.11 & 0.11 & 0.13 \\ 
  Cape Verde & 0.80 & 0.91 & 0.10 & -0.13 & 0.55 & 0.76 & 0.74 & 0.20 & 0.15 & 0.40 & 0.12 & 0.82 & 0.02 & 0.03 & 0.04 & 0.05 & 0.04 & 0.05 \\ 
  Chile & 1.06 & 0.50 & 1.14 & 1.32 & 1.27 & 1.40 & 0.63 & 0.23 & 0.29 & 0.85 & 0.31 & 0.68 & 0.02 & 0.03 & 0.05 & 0.02 & 0.03 & 0.04 \\ 
  China & -1.57 & -0.44 & 0.18 & -0.22 & -0.40 & -0.35 & 0.80 & 0.20 & 0.66 & 0.30 & 0.87 & 0.24 & 0.00 & 0.00 & 0.01 & 0.01 & 0.01 & 0.00 \\ 
  Colombia & -0.15 & -1.39 & -0.11 & 0.23 & -0.52 & -0.28 & 0.67 & 0.13 & 0.64 & 0.80 & 0.13 & 0.83 & 0.01 & 0.02 & 0.01 & 0.01 & 0.01 & 0.02 \\ 
  Costa Rica & 1.05 & 0.74 & 0.30 & 0.52 & 0.56 & 0.72 & 0.33 & 0.14 & 0.20 & 0.80 & 0.12 & 0.77 & 0.06 & 0.05 & 0.05 & 0.06 & 0.04 & 0.07 \\ 
  Croatia & 0.52 & 0.50 & 0.48 & 0.37 & 0.17 & 0.12 & 0.74 & 0.32 & 0.41 & 0.80 & 0.59 & 0.34 & 0.07 & 0.05 & 0.04 & 0.04 & 0.05 & 0.05 \\ 
  Czech Republic & 0.94 & 0.83 & 0.90 & 1.10 & 0.93 & 0.44 & 0.58 & 0.57 & 0.57 & 0.74 & 0.70 & 0.29 & 0.06 & 0.05 & 0.06 & 0.05 & 0.03 & 0.02 \\ 
  Denmark & 1.58 & 1.09 & 1.98 & 1.77 & 1.92 & 2.31 & 0.21 & 0.71 & 0.18 & 0.26 & 0.36 & 0.67 & 0.13 & 0.10 & 0.10 & 0.13 & 0.07 & 0.09 \\ 
  Dominican Republic & 0.12 & 0.02 & -0.44 & -0.15 & -0.57 & -0.75 & 0.63 & 0.29 & 0.63 & 0.44 & 0.13 & 0.52 & 0.03 & 0.04 & 0.03 & 0.03 & 0.03 & 0.03 \\ 
  Ecuador & -0.21 & -0.49 & -0.66 & -0.91 & -0.84 & -0.72 & 0.78 & 0.08 & 0.63 & 0.67 & 0.20 & 0.73 & 0.03 & 0.04 & 0.02 & 0.02 & 0.03 & 0.03 \\ 
  Egypt & -1.15 & -0.84 & -0.46 & -0.49 & -0.24 & -0.65 & 0.70 & 0.25 & 0.45 & 0.80 & 0.07 & 0.04 & 0.01 & 0.01 & 0.01 & 0.01 & 0.02 & 0.02 \\ 
  El Salvador & 0.03 & 0.06 & -0.30 & 0.10 & -0.67 & -0.45 & 0.66 & 0.19 & 0.40 & 0.94 & 0.20 & 0.89 & 0.02 & 0.02 & 0.01 & 0.02 & 0.01 & 0.01 \\ 
  Estonia & 1.09 & 0.70 & 1.04 & 1.44 & 1.10 & 1.15 & 0.48 & 0.56 & 0.31 & 0.65 & 0.81 & 0.17 & 0.19 & 0.09 & 0.04 & 0.12 & 0.04 & 0.04 \\ 
  Ethiopia & -1.19 & -1.50 & -0.69 & -1.03 & -0.71 & -0.54 & 0.69 & 0.20 & 0.64 & 0.55 & 0.26 & 0.46 & 0.01 & 0.01 & 0.01 & 0.01 & 0.01 & 0.01 \\ 
  Fiji & -0.27 & 0.36 & -0.25 & -0.42 & -0.32 & 0.07 & 0.77 & 0.14 & 0.46 & 0.48 & 0.46 & 0.53 & 0.03 & 0.05 & 0.01 & 0.02 & 0.02 & 0.03 \\ 
  Finland & 1.59 & 1.35 & 2.06 & 1.82 & 2.02 & 2.28 & 0.34 & 0.62 & 0.27 & 0.59 & 0.38 & 0.56 & 0.07 & 0.05 & 0.05 & 0.05 & 0.05 & 0.06 \\ 
  France & 1.23 & 0.46 & 1.51 & 1.17 & 1.42 & 1.33 & 0.67 & 0.67 & 0.43 & 0.83 & 0.60 & 0.46 & 0.05 & 0.10 & 0.05 & 0.07 & 0.10 & 0.06 \\ 
  Georgia & -0.04 & -0.64 & 0.08 & 0.33 & -0.26 & 0.01 & 0.65 & 0.41 & 0.55 & 0.85 & 0.39 & 0.32 & 0.03 & 0.02 & 0.02 & 0.01 & 0.05 & 0.02 \\ 
  Germany & 1.38 & 0.89 & 1.59 & 1.58 & 1.66 & 1.84 & 0.39 & 0.64 & 0.64 & 0.67 & 0.79 & 0.39 & 0.13 & 0.10 & 0.07 & 0.08 & 0.12 & 0.06 \\ 
  Ghana & 0.37 & -0.01 & -0.11 & -0.02 & 0.02 & -0.15 & 0.79 & 0.15 & 0.40 & 0.64 & 0.04 & 0.71 & 0.01 & 0.01 & 0.01 & 0.01 & 0.02 & 0.03 \\ 
  Greece & 0.93 & 0.22 & 0.49 & 0.67 & 0.57 & 0.19 & 0.61 & 0.35 & 0.57 & 0.97 & 0.46 & 0.47 & 0.08 & 0.07 & 0.06 & 0.10 & 0.07 & 0.09 \\ 
  Guatemala & -0.34 & -0.71 & -0.67 & -0.22 & -0.98 & -0.76 & 0.95 & 0.06 & 0.37 & 0.98 & 0.14 & 0.83 & 0.02 & 0.02 & 0.01 & 0.02 & 0.01 & 0.01 \\ 
  Honduras & -0.39 & -0.44 & -0.65 & -0.41 & -0.91 & -0.82 & 0.80 & 0.20 & 0.40 & 0.50 & 0.14 & 0.83 & 0.01 & 0.02 & 0.01 & 0.02 & 0.01 & 0.01 \\ 
  Hong Kong & 0.34 & 0.87 & 1.64 & 1.89 & 1.48 & 1.68 & 0.71 & 0.24 & 0.59 & 0.31 & 0.67 & 0.21 & 0.08 & 0.07 & 0.06 & 0.07 & 0.14 & 0.08 \\ 
  Hungary & 0.83 & 0.84 & 0.71 & 0.88 & 0.69 & 0.40 & 0.47 & 0.78 & 0.86 & 0.82 & 0.58 & 0.31 & 0.08 & 0.09 & 0.09 & 0.04 & 0.04 & 0.03 \\ 
  Iceland & 1.45 & 1.35 & 1.72 & 1.37 & 1.79 & 2.03 & 0.33 & 0.60 & 0.14 & 0.52 & 0.30 & 0.64 & 0.11 & 0.07 & 0.13 & 0.09 & 0.08 & 0.09 \\ 
  India & 0.36 & -1.02 & 0.05 & -0.27 & 0.07 & -0.35 & 0.77 & 0.48 & 0.56 & 0.40 & 0.51 & 0.26 & 0.01 & 0.02 & 0.01 & 0.01 & 0.01 & 0.01 \\ 
  Indonesia & -0.03 & -1.10 & -0.17 & -0.28 & -0.59 & -0.68 & 0.78 & 0.14 & 0.46 & 0.48 & 0.62 & 0.38 & 0.01 & 0.01 & 0.01 & 0.01 & 0.01 & 0.01 \\ 
  Iran & -1.36 & -1.15 & -0.56 & -1.47 & -0.83 & -0.70 & 0.56 & 0.40 & 0.42 & 0.57 & 0.14 & 0.39 & 0.02 & 0.02 & 0.02 & 0.02 & 0.01 & 0.02 \\ 
  Iraq & -1.29 & -2.29 & -1.48 & -1.48 & -1.57 & -1.36 & 0.95 & 0.30 & 0.70 & 0.85 & 0.25 & 0.17 & 0.02 & 0.01 & 0.02 & 0.02 & 0.01 & 0.02 \\ 
  Ireland & 1.42 & 1.16 & 1.56 & 1.65 & 1.64 & 1.60 & 0.32 & 0.69 & 0.66 & 0.38 & 0.28 & 0.62 & 0.12 & 0.09 & 0.07 & 0.12 & 0.11 & 0.10 \\ 
  Israel & 0.66 & -1.11 & 1.21 & 1.15 & 0.98 & 0.83 & 0.27 & 0.50 & 0.46 & 0.78 & 0.39 & 0.28 & 0.27 & 0.11 & 0.07 & 0.10 & 0.13 & 0.10 \\ 
  Italy & 1.03 & 0.54 & 0.54 & 0.79 & 0.48 & 0.41 & 0.52 & 0.73 & 0.68 & 0.75 & 0.60 & 0.30 & 0.08 & 0.12 & 0.06 & 0.05 & 0.07 & 0.06 \\ 
  Jamaica & 0.56 & -0.03 & 0.17 & 0.17 & -0.35 & -0.15 & 0.45 & 0.39 & 0.68 & 0.13 & 0.13 & 0.83 & 0.01 & 0.04 & 0.01 & 0.03 & 0.02 & 0.02 \\ 
  Japan & 1.00 & 1.04 & 1.48 & 1.14 & 1.38 & 1.42 & 0.54 & 0.46 & 0.94 & 0.91 & 0.88 & 0.42 & 0.03 & 0.03 & 0.06 & 0.06 & 0.04 & 0.02 \\ 
  Jordan & -0.62 & -0.27 & 0.07 & 0.18 & 0.31 & 0.14 & 0.52 & 0.22 & 0.34 & 0.48 & 0.13 & 0.30 & 0.17 & 0.07 & 0.11 & 0.16 & 0.07 & 0.12 \\ 
  Kazakhstan & -1.10 & 0.10 & -0.33 & -0.22 & -0.68 & -0.88 & 0.86 & 0.22 & 0.46 & 0.87 & 0.82 & 0.21 & 0.04 & 0.07 & 0.06 & 0.05 & 0.06 & 0.04 \\ 
  Kenya & -0.35 & -1.15 & -0.49 & -0.28 & -0.75 & -0.97 & 0.69 & 0.25 & 0.59 & 0.49 & 0.25 & 0.57 & 0.01 & 0.01 & 0.01 & 0.01 & 0.01 & 0.02 \\ 
  Kuwait & -0.53 & 0.30 & 0.01 & 0.14 & 0.41 & 0.22 & 0.77 & 0.29 & 0.46 & 0.61 & 0.34 & 0.28 & 0.11 & 0.12 & 0.09 & 0.17 & 0.15 & 0.10 \\ 
  Latvia & 0.81 & 0.53 & 0.72 & 0.99 & 0.67 & 0.35 & 0.50 & 0.65 & 0.12 & 0.67 & 0.71 & 0.14 & 0.17 & 0.13 & 0.09 & 0.11 & 0.05 & 0.04 \\ 
  Lebanon & -0.39 & -1.28 & -0.46 & -0.28 & -0.59 & -0.80 & 0.68 & 0.35 & 0.57 & 0.47 & 0.14 & 0.23 & 0.08 & 0.05 & 0.07 & 0.08 & 0.05 & 0.04 \\ 
  Libya & -1.64 & -0.90 & -1.38 & -1.80 & -1.31 & -1.27 & 0.75 & 0.35 & 0.49 & 0.64 & 0.22 & 0.32 & 0.06 & 0.05 & 0.04 & 0.06 & 0.05 & 0.05 \\ 
  Lithuania & 0.93 & 0.71 & 0.78 & 1.02 & 0.74 & 0.50 & 0.44 & 0.59 & 0.20 & 0.66 & 0.82 & 0.16 & 0.11 & 0.06 & 0.04 & 0.06 & 0.03 & 0.03 \\ 
  Luxembourg & 1.54 & 1.41 & 1.81 & 1.75 & 1.84 & 1.98 & 0.48 & 0.56 & 0.48 & 0.75 & 0.59 & 0.50 & 0.13 & 0.15 & 0.10 & 0.13 & 0.15 & 0.11 \\ 
  Malawi & -0.18 & -0.08 & -0.60 & -0.56 & -0.26 & -0.53 & 0.70 & 0.30 & 0.40 & 0.50 & 0.25 & 0.61 & 0.02 & 0.02 & 0.01 & 0.01 & 0.02 & 0.02 \\ 
  Malaysia & -0.32 & 0.24 & 1.05 & 0.66 & 0.48 & 0.22 & 0.97 & 0.25 & 0.49 & 0.37 & 0.42 & 0.55 & 0.07 & 0.04 & 0.03 & 0.05 & 0.05 & 0.07 \\ 
  Malta & 1.18 & 1.25 & 1.00 & 1.21 & 1.26 & 0.77 & 0.55 & 0.59 & 0.48 & 0.91 & 0.46 & 0.64 & 0.07 & 0.09 & 0.05 & 0.14 & 0.06 & 0.07 \\ 
  Mexico & 0.10 & -0.60 & 0.11 & 0.23 & -0.48 & -0.50 & 0.81 & 0.30 & 0.69 & 0.82 & 0.24 & 0.97 & 0.03 & 0.05 & 0.01 & 0.03 & 0.01 & 0.03 \\ 
  Moldova & -0.15 & -0.39 & -0.61 & -0.17 & -0.39 & -0.70 & 0.90 & 0.27 & 0.39 & 0.95 & 0.71 & 0.19 & 0.02 & 0.02 & 0.02 & 0.01 & 0.03 & 0.01 \\ 
  Morocco & -0.61 & -0.33 & -0.11 & -0.16 & -0.08 & -0.24 & 0.70 & 0.46 & 0.53 & 0.68 & 0.14 & 0.25 & 0.01 & 0.01 & 0.01 & 0.01 & 0.02 & 0.01 \\ 
  Mozambique & -0.24 & -0.26 & -0.60 & -0.49 & -0.74 & -0.58 & 0.84 & 0.15 & 0.38 & 0.44 & 0.11 & 0.79 & 0.02 & 0.02 & 0.01 & 0.01 & 0.02 & 0.02 \\ 
  Namibia & 0.43 & 0.52 & 0.11 & 0.06 & 0.16 & 0.35 & 0.64 & 0.30 & 0.39 & 0.45 & 0.35 & 0.61 & 0.04 & 0.05 & 0.04 & 0.04 & 0.05 & 0.04 \\ 
  Nepal & -0.48 & -1.21 & -0.84 & -0.65 & -0.64 & -0.65 & 0.65 & 0.30 & 0.40 & 0.40 & 0.53 & 0.30 & 0.02 & 0.03 & 0.02 & 0.01 & 0.01 & 0.01 \\ 
  Netherlands & 1.55 & 1.09 & 1.90 & 1.81 & 1.81 & 2.04 & 0.41 & 0.76 & 0.18 & 0.54 & 0.64 & 0.65 & 0.11 & 0.13 & 0.12 & 0.08 & 0.11 & 0.10 \\ 
  New Zealand & 1.59 & 1.37 & 1.74 & 1.80 & 1.89 & 2.24 & 0.28 & 0.73 & 0.56 & 0.47 & 0.36 & 0.68 & 0.16 & 0.15 & 0.06 & 0.08 & 0.12 & 0.12 \\ 
  Nigeria & -0.62 & -1.82 & -1.00 & -0.81 & -1.08 & -1.12 & 0.80 & 0.30 & 0.60 & 0.55 & 0.13 & 0.84 & 0.01 & 0.01 & 0.01 & 0.01 & 0.01 & 0.01 \\ 
  North Macedonia & -0.09 & -0.50 & -0.18 & 0.18 & -0.28 & -0.36 & 0.90 & 0.22 & 0.46 & 0.86 & 0.62 & 0.34 & 0.03 & 0.02 & 0.06 & 0.03 & 0.03 & 0.04 \\ 
  Norway & 1.63 & 1.31 & 1.89 & 1.49 & 1.94 & 2.13 & 0.33 & 0.66 & 0.12 & 0.51 & 0.36 & 0.53 & 0.11 & 0.09 & 0.13 & 0.08 & 0.08 & 0.07 \\ 
  Pakistan & -0.92 & -1.97 & -0.61 & -0.67 & -0.80 & -0.93 & 0.55 & 0.14 & 0.50 & 0.69 & 0.50 & 0.26 & 0.02 & 0.04 & 0.01 & 0.03 & 0.01 & 0.00 \\ 
  Panama & 0.55 & 0.12 & 0.17 & 0.36 & -0.13 & -0.38 & 0.93 & 0.12 & 0.44 & 0.85 & 0.13 & 0.82 & 0.06 & 0.06 & 0.04 & 0.05 & 0.06 & 0.06 \\ 
  Paraguay & -0.20 & -0.48 & -0.86 & -0.50 & -0.79 & -1.05 & 0.70 & 0.13 & 0.40 & 0.85 & 0.20 & 0.56 & 0.02 & 0.05 & 0.02 & 0.02 & 0.04 & 0.02 \\ 
  Peru & 0.03 & -0.73 & -0.28 & 0.39 & -0.55 & -0.40 & 0.64 & 0.16 & 0.42 & 0.87 & 0.25 & 0.46 & 0.01 & 0.03 & 0.02 & 0.02 & 0.02 & 0.03 \\ 
  Philippines & 0.05 & -1.17 & -0.01 & -0.06 & -0.43 & -0.56 & 0.94 & 0.32 & 0.64 & 0.44 & 0.27 & 0.42 & 0.02 & 0.02 & 0.01 & 0.01 & 0.02 & 0.01 \\ 
  Poland & 0.94 & 0.63 & 0.56 & 0.89 & 0.64 & 0.57 & 0.68 & 0.60 & 0.63 & 0.93 & 0.39 & 0.29 & 0.04 & 0.04 & 0.04 & 0.03 & 0.06 & 0.02 \\ 
  Portugal & 1.26 & 1.02 & 1.06 & 0.95 & 1.16 & 1.00 & 0.63 & 0.28 & 0.31 & 0.96 & 0.29 & 0.35 & 0.05 & 0.07 & 0.05 & 0.10 & 0.07 & 0.10 \\ 
  Qatar & -0.92 & 1.00 & 0.71 & 0.48 & 0.74 & 0.87 & 0.78 & 0.28 & 0.50 & 0.59 & 0.36 & 0.29 & 0.12 & 0.11 & 0.09 & 0.17 & 0.15 & 0.10 \\ 
  Romania & 0.48 & 0.15 & -0.19 & 0.35 & 0.06 & -0.23 & 0.90 & 0.30 & 0.42 & 0.90 & 0.52 & 0.20 & 0.03 & 0.03 & 0.02 & 0.02 & 0.02 & 0.02 \\ 
  Russia & -0.81 & -1.07 & -0.37 & -0.40 & -0.86 & -0.95 & 0.91 & 0.38 & 0.36 & 0.93 & 0.80 & 0.20 & 0.04 & 0.04 & 0.04 & 0.03 & 0.04 & 0.02 \\ 
  Saudi Arabia & -1.68 & -0.31 & -0.05 & 0.09 & 0.07 & 0.01 & 0.85 & 0.25 & 0.54 & 0.69 & 0.36 & 0.43 & 0.11 & 0.07 & 0.08 & 0.14 & 0.09 & 0.12 \\ 
  Senegal & 0.10 & -0.27 & -0.27 & -0.20 & -0.15 & -0.13 & 0.69 & 0.25 & 0.44 & 0.54 & 0.25 & 0.45 & 0.01 & 0.02 & 0.01 & 0.02 & 0.02 & 0.02 \\ 
  Serbia & -0.09 & -0.54 & -0.22 & -0.24 & -0.57 & -0.52 & 0.85 & 0.25 & 0.43 & 0.91 & 0.53 & 0.29 & 0.04 & 0.03 & 0.02 & 0.03 & 0.05 & 0.04 \\ 
  Sierra Leone & -0.48 & -0.59 & -1.24 & -0.98 & -1.01 & -0.78 & 0.69 & 0.20 & 0.39 & 0.49 & 0.16 & 0.59 & 0.01 & 0.01 & 0.01 & 0.01 & 0.01 & 0.01 \\ 
  Singapore & -0.06 & 1.25 & 2.19 & 2.03 & 1.67 & 2.16 & 0.78 & 0.22 & 0.50 & 0.21 & 0.63 & 0.44 & 0.12 & 0.08 & 0.06 & 0.18 & 0.15 & 0.11 \\ 
  Slovakia & 0.90 & 0.82 & 0.73 & 0.86 & 0.53 & 0.31 & 0.99 & 0.52 & 0.99 & 0.52 & 0.77 & 0.28 & 0.07 & 0.03 & 0.08 & 0.05 & 0.03 & 0.02 \\ 
  Slovenia & 1.03 & 0.90 & 0.96 & 0.78 & 1.00 & 0.86 & 0.72 & 0.27 & 0.22 & 0.87 & 0.51 & 0.47 & 0.07 & 0.05 & 0.10 & 0.04 & 0.06 & 0.05 \\ 
  South Africa & 0.67 & -0.17 & 0.47 & 0.40 & 0.09 & 0.28 & 0.49 & 0.63 & 0.62 & 0.48 & 0.33 & 0.62 & 0.04 & 0.06 & 0.03 & 0.03 & 0.04 & 0.04 \\ 
  South Korea & 0.71 & 0.39 & 1.08 & 0.90 & 1.00 & 0.50 & 0.60 & 0.18 & 0.39 & 0.84 & 0.99 & 0.29 & 0.02 & 0.03 & 0.03 & 0.06 & 0.05 & 0.02 \\ 
  Spain & 1.12 & 0.20 & 1.26 & 1.07 & 1.11 & 1.06 & 0.58 & 0.49 & 0.43 & 0.84 & 0.47 & 0.44 & 0.06 & 0.08 & 0.05 & 0.07 & 0.08 & 0.08 \\ 
  Sri Lanka & -0.25 & -0.79 & -0.18 & -0.13 & 0.04 & -0.32 & 0.80 & 0.35 & 0.10 & 0.45 & 0.45 & 0.39 & 0.01 & 0.01 & 0.02 & 0.01 & 0.01 & 0.01 \\ 
  Suriname & 0.36 & 0.24 & -0.23 & -0.63 & -0.10 & -0.06 & 0.82 & 0.45 & 0.37 & 0.88 & 0.19 & 0.79 & 0.05 & 0.05 & 0.05 & 0.08 & 0.10 & 0.09 \\ 
  Sweden & 1.55 & 1.15 & 1.85 & 1.62 & 1.88 & 2.17 & 0.35 & 0.67 & 0.10 & 0.34 & 0.51 & 0.72 & 0.13 & 0.12 & 0.16 & 0.15 & 0.08 & 0.14 \\ 
  Switzerland & 1.54 & 1.31 & 1.96 & 1.67 & 1.88 & 2.08 & 0.40 & 0.63 & 0.65 & 0.61 & 0.68 & 0.59 & 0.15 & 0.14 & 0.11 & 0.11 & 0.13 & 0.13 \\ 
  Syria & -1.74 & -1.43 & -1.20 & -1.33 & -1.03 & -1.25 & 0.78 & 0.34 & 0.51 & 0.59 & 0.29 & 0.31 & 0.04 & 0.02 & 0.03 & 0.04 & 0.03 & 0.03 \\ 
  São Tomé and Príncipe & 0.26 & 0.49 & -0.67 & -0.83 & -0.53 & -0.10 & 0.75 & 0.37 & 0.24 & 0.69 & 0.32 & 0.41 & 0.01 & 0.02 & 0.01 & 0.03 & 0.02 & 0.05 \\ 
  Tanzania & -0.38 & -0.42 & -0.56 & -0.44 & -0.42 & -0.60 & 0.69 & 0.25 & 0.40 & 0.49 & 0.34 & 0.38 & 0.01 & 0.01 & 0.01 & 0.01 & 0.01 & 0.01 \\ 
  Thailand & -0.38 & -0.70 & 0.29 & 0.32 & 0.08 & -0.33 & 0.62 & 0.19 & 0.33 & 0.61 & 0.31 & 0.43 & 0.03 & 0.01 & 0.02 & 0.03 & 0.02 & 0.02 \\ 
  Trinidad and Tobago & 0.57 & 0.09 & 0.26 & 0.36 & 0.01 & -0.07 & 0.46 & 0.16 & 0.57 & 0.54 & 0.13 & 0.78 & 0.03 & 0.06 & 0.02 & 0.03 & 0.04 & 0.04 \\ 
  Tunisia & -0.53 & -0.26 & 0.15 & -0.18 & -0.02 & -0.18 & 0.70 & 0.40 & 0.40 & 0.75 & 0.44 & 0.35 & 0.01 & 0.01 & 0.01 & 0.01 & 0.01 & 0.01 \\ 
  Turkey & -0.32 & -1.00 & 0.13 & 0.24 & -0.04 & -0.14 & 0.66 & 0.37 & 0.45 & 0.84 & 0.46 & 0.48 & 0.03 & 0.03 & 0.02 & 0.04 & 0.04 & 0.04 \\ 
  Ukraine & -0.20 & -0.76 & -0.60 & -0.49 & -0.84 & -0.93 & 0.91 & 0.26 & 0.28 & 0.94 & 0.84 & 0.15 & 0.03 & 0.04 & 0.04 & 0.04 & 0.05 & 0.04 \\ 
  United Arab Emirates & -0.89 & 0.81 & 1.05 & 0.78 & 0.59 & 0.84 & 0.74 & 0.31 & 0.50 & 0.56 & 0.39 & 0.26 & 0.11 & 0.13 & 0.08 & 0.17 & 0.14 & 0.11 \\ 
  United Kingdom & 1.32 & 0.51 & 1.67 & 1.71 & 1.67 & 1.84 & 0.37 & 0.84 & 0.64 & 0.37 & 0.50 & 0.66 & 0.10 & 0.14 & 0.06 & 0.10 & 0.07 & 0.10 \\ 
  United States & 1.14 & 0.42 & 1.54 & 1.46 & 1.55 & 1.39 & 0.44 & 0.82 & 0.61 & 0.48 & 0.27 & 0.66 & 0.12 & 0.20 & 0.05 & 0.11 & 0.10 & 0.11 \\ 
  Uruguay & 1.10 & 0.92 & 0.58 & 0.50 & 0.64 & 1.20 & 0.61 & 0.36 & 0.38 & 0.98 & 0.26 & 0.53 & 0.02 & 0.03 & 0.03 & 0.03 & 0.03 & 0.03 \\ 
  Venezuela & -0.83 & -1.15 & -1.16 & -1.47 & -1.63 & -1.18 & 0.80 & 0.12 & 0.72 & 0.76 & 0.16 & 0.99 & 0.04 & 0.04 & 0.04 & 0.02 & 0.03 & 0.07 \\ 
  Vietnam & -1.38 & 0.21 & -0.13 & -0.51 & -0.35 & -0.54 & 0.70 & 0.20 & 0.40 & 0.30 & 0.57 & 0.35 & 0.00 & 0.01 & 0.00 & 0.01 & 0.00 & 0.00 \\ 
  Zambia & -0.32 & 0.14 & -0.80 & -0.51 & -0.47 & -0.57 & 0.60 & 0.35 & 0.40 & 0.50 & 0.30 & 0.42 & 0.02 & 0.02 & 0.02 & 0.01 & 0.02 & 0.03 \\ 
\end{longtable}
\end{landscape}

\end{document}